\documentclass[a4paper,10pt,twoside]{cpc-hepnp}
\usepackage{CJK,upgreek,fancyhdr}
\usepackage{multicol}
\usepackage{graphicx}
\usepackage{booktabs}
\usepackage{amssymb,bm,mathrsfs,bbm,amscd}
\usepackage[tbtags]{amsmath}
\usepackage{lastpage}
\usepackage{color}

\usepackage{epstopdf}
\usepackage{hyperref}
\hypersetup{colorlinks=true, linkcolor=blue, urlcolor=blue, citecolor=black, bookmarks=false, pdfstartview={FitH}}

\begin{document}
\begin{CJK*}{GB}{gbsn}

\fancyhead[c]{\small Chinese Physics C~~~Vol. xx, No. x (201x) xxxxxx}
\fancyfoot[C]{\small 010201-\thepage}

\footnotetext[0]{Received 31 June 2015}

\title{Fermion scattering by a class of Bardeen black holes}

\author{ Ciprian A. Sporea$^{1;1)}$\email{ciprian.sporea@e-uvt.ro} }

\maketitle

\address{ $^1$ West University of Timi\c soara, V.  P\^ arvan Ave.  4, RO-300223 Timi\c soara, Romania }

\begin{abstract}
In this paper the scattering of fermions by a class of Bardeen black holes is investigated. After obtaining the scattering modes by solving the Dirac equation in this geometry, we use the partial wave method to derive an analytical expression for the phase shifts that enter into the definitions of partial amplitudes that define the scattering cross sections and the induced polarization. It is then shown that, like in the case of Schwarzschild and Reissner-Nordstr\"om, the phenomena of glory and spiral scattering are present.
\end{abstract}

\begin{keyword}
Bardeen black hole, fermion scattering, partial wave method.
\end{keyword}

\begin{pacs}
04.70.-s, 03.65.Nk, 04.62.+v.
\end{pacs}

\footnotetext[0]{\hspace*{-3mm}\raisebox{0.3ex}{$\scriptstyle\copyright$}2013
Chinese Physical Society and the Institute of High Energy Physics
of the Chinese Academy of Sciences and the Institute
of Modern Physics of the Chinese Academy of Sciences and IOP Publishing Ltd}%

\begin{multicols}{2}

\section{Introduction}

The existence of black holes is well motivated theoretically in General Relativity or in other modified theories of gravity. Thus far, black holes have not been directly detected or observed. However, there exist plenty of indirect evidences that indicate that such objects are real and do exist in nature. As shown by Penrose and Hawking \cite{hawking.book} the occurrence of singularities is inevitable in General Relativity. This brings many issues to the table, like the black hole information paradox \cite{SW1, SW2, SW3} or the more recently discovered paradox of black hole's "firewalls" \cite{firewall, fire1, fire2, fire3, fire4}. These paradoxes emerge because of the incompatibility between quantum theory and General Relativity (GR). It is widely believed that in a quantum theory of gravity the singularities contained in black holes will be removed. However, even in the early stages of investigations on singularities \cite{sakharov,gliner} in GR, there have been proposal of black hole models that could avoid the occurrence of a singularity. These black holes are said to be "regular" in the sense of being singularity-free.

The first proposal of a regular black hole solution was made by Bardeen in Ref. \cite{Bardeen1}, and since then many other models of spherically symmetric regular black holes where presented in the literature \cite{borde,rbh1,rbh2,rbh3,rbh4,rbh5,rbh6,rbh7,rbh8}. In Ref. \cite{Ayon} the authors have shown that the Bardeen black hole model can be physically interpreted as the gravitational field produced by a nonlinear magnetic monopole. Later on, this interpretation was extended to include also nonlinear electric charges so that one can now say that regular black holes models can have as a source a nonlinear electromagnetic field. More recently, in Ref. \cite{wang} the authors extended the Bardeen solution to an entire class of Bardeen-like black holes that can be regular or not.

In this paper we will study the scattering of fermions (spin $1/2$) by a Bardeen regular black hole and by a Bardeen-class of black holes (as constructed in \cite{wang}). We will use the partial wave method to obtain analytical expressions for the phase shifts that enter into the definition of partial amplitudes defining the scattering cross sections and the induced polarization. To our knowledge this is the first study to report analytical phase shifts for spin $1/2$ wave scattering by regular black holes. In previous works only the absorbtion of fermions by Bardeen black holes was investigated numerically in \cite{huang2}, while in Refs. \cite{huang1, Macedo2, Macedo2a, Macedo2b} the case of massless scalar scattering by regular black holes was treated also numerically. Studies dedicated to fermion scattering by other types of spherically symmetric black holes can be found for example in Refs. \cite{unruh, Das, Jin, Doran, Jung, Gaina, Gaina1, ChaoLin, dolan, Rogatko, Liao, Ghosh, sporea1, sporea2, sporea3}.

Both Bardeen regular black holes and the Bardeen-class type of black holes possess nonlinear magnetic (monopole) charges \cite{Ayon, wang}. This implies the existence of an electromagnetic potential of the form $A=Q_m\cos\theta\,d\phi$, with $Q_m$ the total magnetic charge. In this work we will neglect the interaction between this potential and the charge of the fermion and focus instead only on studying the scattering resulting from the "pure" gravitational interaction between the fermion and the black hole. However, even in this approximation the black hole magnetic monopole charge will still influence the scattering patterns through the presence of $Q_m$ into the metric function and into the resulting scattering modes of the radial Dirac equation.

The rest of the paper is structured as follows. In Section \ref{section2} the Bardeen-class of black holes are presented very briefly. Section \ref{section3} starts with a very short review of the Dirac equation in spherically symmetric black hole geometries and continues with the search for scattering modes in the Bardeen spacetime. In the last part of this section the main result of the paper is presented, namely the form of the analytical phase shifts resultant from applying the partial wave method on the scattering modes derived earlier. The next Section \ref{sec.results} is dedicated to a graphical analysis of the induced polarization and the scattering cross sections in which the presence of a backward "glory" and "spiral scattering" (orbiting) oscillations are shown to be present. The main conclusions and some final remarks are given in the last Section \ref{sec.final} of the paper.

\section{A class of Bardeen-like black holes}\label{section2}

In Ref. \cite{wang} a class of spherically symmetric and asymptotically flat Bardenn-like black holes depending on two-parameters was constructed, having the following line element
\begin{equation}\label{ds2}
\begin{split}
& ds^2=h(r) dt^2-\frac{dr^2}{h(r)}-r^2\left( d\theta^2+\sin^2\theta d\phi^2 \right) \\
& h(r)=1-\frac{2M_s}{r}-\frac{2Mr^{\gamma-1}}{\sqrt{(r^2+Q^2)^\gamma}}
\end{split}
\end{equation}
These black holes have a singularity if the parameter $M_s$, referred from now on as the "Schwarzschild mass", takes a nonzero value. Otherwise, if $M_s=0$, regular black holes are obtained and the Bardeen black hole corresponds to the particular choice $M_s=0$ and $\gamma=3$. The term $M$ can be interpreted as the mass of the nonlinear magnetic monopole. Moreover, the sum
\begin{equation}\label{ADM}
M_{ADM}=M_s + M
\end{equation}
constitutes the ADM mass of the black hole obtained from the asymptotic form of the metric function $h(r)$. The parameter $Q$ entering eq. (\ref{ds2}) is related to the magnetic monopole charge $Q_m$ by the relation $Q=2Q_m^2/M$.

As shown in Ref. \cite{wang} the Lagrangian density, for which eq. (\ref{ds2}) is a solution of the coupled Einstein-Maxwell filed equations, is given by the expression
\begin{equation}
\mathcal{L}=\frac{4\gamma}{\alpha}\frac{(\alpha F_{\mu\nu}F^{\mu\nu})^{5/4}}{(1+\sqrt{\alpha F_{\mu\nu}F^{\mu\nu}})^{1+\gamma/2}}
\end{equation}
where $\alpha$ has dimension of length squared and $\gamma$ is a dimensionless constant. In the weak field limit one gets a vector field that is slightly stronger when compared with a Maxwell field.

\section{Dirac fermions and scattering cross sections}\label{section3}

\subsection{Dirac equation. Preliminaries}

The Dirac equation
\begin{equation}\label{de13}
i\gamma^a D_a\psi -m\psi = 0
\end{equation}
can be brought to the following explicit form \cite{cota1}
\begin{equation}\label{ecd}
\left( i\gamma^a e^\mu_a\partial_\mu -m \right)\psi + \frac{i}{2}\frac{1}{\sqrt{-g}}\partial_\mu(\sqrt{-g}\,e^\mu_a)\gamma^a\psi -\frac{1}{4}\{\gamma^a , S^{\,b}_{\,\,\,c} \}\omega^{\,c}_{\, a b}\psi = 0
\end{equation}
where $g=\det (g_{\mu\nu})$ and the covariant derivative is defined by $D_a = \partial_a + \frac{i}{2}S^{\,b}_{\,\,c}\,\omega^{\,c}_{\, a b}$, with $S^{a b}=\frac{i}{4}[\gamma^a,\gamma^b]$ the generators of the $SL(2,\mathbb{C})$ group; $\gamma^a$ the point-independent Dirac matrices obeying $\{\gamma^a,\gamma^b\}=2\eta^{a b}$ and $\omega^{\,c}_{\, a b}$ is a spin-connection \begin{equation}\label{de12}
\omega^{\,c}_{\, ab} = e^\mu_a e^\nu_b \left( \hat e^c_\lambda\Gamma^\lambda_{\mu\nu} - \hat e^c_{\nu,\mu} \right)
\end{equation}
with $\Gamma^\lambda_{\mu\nu}$ the usual Christoffel symbols. The tetrad fields $e_a(x)$ and $\hat e^a(x)$ are point dependent defining (non-holonomic) local frames and co-frames and the following relations hold
\begin{equation}\label{de5}
\begin{split}
& e_a=e_a^\mu\partial_\mu , \qquad \hat e^a=\hat e^a_\mu dx^\mu \\
& e_a(x) e_b(x) = \eta_{a b}, \qquad \hat e^a(x) \hat e^b(x) = \eta^{a b} \\
& ds^2=\eta_{a b}\hat e^a_\mu dx^\mu \hat e^b_\nu dx^\nu = g_{\mu\nu}(x)dx^\mu dx^\nu
\end{split}
\end{equation}

Introducing now the so called Cartesian gauge \cite{Villalba, cota1, cota2}, that for a spherically symmetric line element of the form (\ref{ds2}) is defined by the following tetrad fields:
\begin{equation}\label{de38}
\openup 2\jot
\begin{split}
&\hat e^0 = \sqrt{h}\,dt \\
&\hat e^1 = \frac{1}{\sqrt{h}}\sin\theta\cos\phi\,dr + r\,\cos\theta\cos\phi\,d\theta - r\,\sin\theta\sin\phi\,d\phi\\
&\hat e^2 = \frac{1}{\sqrt{h}}\sin\theta\sin\phi\,dr + r\,\cos\theta\sin\phi\,d\theta + r\,\sin\theta\cos\phi\,d\phi\\
&\hat e^3 =\frac{1}{\sqrt{h}}\cos\theta\,dr - r\,\sin\theta\,d\theta ,
\end{split}
\end{equation}
the Dirac equation (\ref{ecd}) can be reduced to only a radial equation. The angular part of the Dirac equation is the same as in the Dirac theory from flat spacetime and its solutions are the usual 4-component angular spinors $\Phi^\pm_{m,\kappa}(\theta,\phi)$ \cite{Thaller, Landau}. This is due to the fact that in the Cartesian gauge (\ref{de38}) the Dirac equation is manifestly covariant under rotations \cite{cota2}. Using this gauge it was possible to find complete analytical solutions to the Dirac equation on de Sitter/anti-de Sitter space time \cite{cota3, cota3a, crucean1, crucean2} and approximative analytical solutions in black hole geometries \cite{cota, sporea4, sporea1, sporea2, sporea3} that were later on used to study different aspects of the scattering problem on those spacetimes \cite{sporea1, sporea2, sporea3, crucean3, crucean4, crucean5, crucean6}.

The remaining unsolved radial part of the Dirac equation is found by assuming the following type of particle-like solutions with a given energy $E$
\begin{equation}\label{rad1}
\begin{split}
\psi(x)=&\psi_{E,j,m,\kappa}(t,r,\theta,\phi) =\\
&\frac{e^{-iEt}}{r\,h(r)^{1/4}}\left\{ f^+_{E,\kappa}(r)\Phi^+_{m,\kappa}(\theta, \phi) + f^-_{E,\kappa}(r)\Phi^-_{m,\kappa}(\theta, \phi) \right\}
\end{split}
\end{equation}
with $f^\pm_{E,\kappa}(r)$ two unknown radial wave functions. As in Refs. \cite{cota, sporea1} the radial Dirac equation can be put into a matrix form

\begin{equation}\label{rad6}
\renewcommand{\arraystretch}{1.8}
\begin{split}
& \left(\begin{array}{cc}
m\sqrt{h(r)}& -h(r)\frac{\textstyle d}{\textstyle dr}+\frac{\textstyle \kappa}{\textstyle r}\sqrt{h(r)}\\
h(r)\frac{\textstyle d}{\textstyle dr}+\frac{\textstyle \kappa}{\textstyle r}\sqrt{h(r)}& -m\sqrt{h(r)}
\end{array}\right)\times \\
&\qquad \times\left(\begin{array}{c}
f^+_{E,\kappa}(r)  \\
f^-_{E,\kappa}(r)
\end{array}\right) = E \left(\begin{array}{c}
f^+_{E,\kappa}(r)  \\
f^-_{E,\kappa}(r)
\end{array}\right)
\end{split}
\end{equation}

\subsection{Scattering modes}\label{sec.scatt}

Inserting the line element (\ref{ds2}) into eq. (\ref{rad6}) we obtain a system of two differential equations that has no analytical solutions due to the complex form of the line element (\ref{ds2}). However, because we are interested here only in finding the scattering modes, one can approximate eq. (\ref{rad6}) in the asymptotic region of the Bardeen black hole specified by the line element in eq. (\ref{ds2}) and find approximative analytical solutions. On these new solutions a partial wave method will be used in order to compute the (elastic) scattering cross section and the induced polarization that results after the interaction of a fermion beam with the black hole.

Let us start by introducing the new variable
\begin{equation}\label{sol1}
x=\sqrt{\frac{z}{r_+}-1}, \qquad z=\sqrt{r^2+Q^2},
\end{equation}
with $r_+$ the radius of the black hole horizon. In terms of the new introduced variable the function $h(r)$ becomes
\begin{equation}
h=1-\frac{2M_s}{r_+}\frac{1}{\sqrt{(1+x^2)^2-\delta}}-\frac{2M}{r_+}\frac{1}{1+x^2}\left[1-\frac{\delta}{(1+x^2)^2} \right]^{\frac{\gamma-1}{2}}
\end{equation}
and where the notation $\delta=\left( Q/r_+ \right)^2$ was introduced. The radial Dirac equation (\ref{rad6}) is in fact a system of two differential equations for the radial wave functions $f^\pm(x)$ that, after multiplying each equation by $x/[r_+(1+x^2)]$ and expressing all the terms as a function of the new variable $x$ given by eq. (\ref{sol1}), is equivalent to eq. (\ref{sol4})

\begin{equation}\label{sol4}
\openup 2\jot
\begin{split}
& \left[ \frac{1}{1+x^2}\sqrt{1-\frac{\delta}{(1+x^2)^2}}\left\{ \frac{\,}{\,}... \right\}\frac{1}{2}\frac{d}{dx} + \frac{\kappa}{\sqrt{(1+x^2)^2-\delta}} \times \right. \\
&\times \left.\left\{ \frac{\,}{\,}... \right\}^{\frac{1}{2}} \right] f^+_{E,\kappa}(x) - \left[ \mu \left\{ \frac{\,}{\,}... \right\}^{\frac{1}{2}} +\epsilon \left(x+\frac{1}{x} \right) \right]f^-_{E,\kappa}(x) = 0 \\
& \left[ \frac{1}{1+x^2}\sqrt{1-\frac{\delta}{(1+x^2)^2}}\left\{ \frac{\,}{\,}... \right\}\frac{1}{2}\frac{d}{dx} - \frac{\kappa}{\sqrt{(1+x^2)^2-\delta}} \times \right. \\
&\times \left.\left\{ \frac{\,}{\,}... \right\}^{\frac{1}{2}} \right] f^-_{E,\kappa}(x) - \left[ \mu \left\{ \frac{\,}{\,}... \right\}^{\frac{1}{2}} -\epsilon \left(x+\frac{1}{x} \right) \right]f^+_{E,\kappa}(x) = 0 \\
& where\,\,\,\left\{ \frac{\,}{\,}... \right\}= \frac{(1+x^2)^2}{x^2}-\frac{2M_s}{r_+}\frac{(1+x^2)^2}{x^2\sqrt{(1+x^2)^2-\delta}}  \\
&-\frac{2M}{r_+}\frac{1+x^2}{x^2} \left[1-\frac{\delta}{(1+x^2)^2}\right]^{\frac{\gamma-1}{2}}, \qquad \mu=r_+ m, \quad \epsilon=r_+ E
\end{split}
\end{equation}

The system of differential equations obtained in eq. (\ref{sol4}) can not be solved analytically as it is. However, if we are interested in obtaining an analytical solution this can be done if we restrict to a domain far away from the black hole event horizon. One obtains a more simple system of differential equations that have analytical solutions. By making a Taylor expansion with respect to $1/x$ and discarding the terms of the order $O(1/x^2)$ and higher, while keeping all the other remaining terms, the system of equations, valid in the asymptotic region of the black hole for the two radial wave function, reduces to
\begin{equation}\label{sol7}
\begin{split}
& \left( \frac{1}{2}\frac{d}{dx}+\frac{\kappa}{x} \right)f^+_{E,\kappa}(x) - x(\epsilon+\mu)f^-_{E,\kappa}(x) \\
&  - \frac{1}{x}\left[\epsilon+ \mu\left(1-\frac{M_{ADM}}{r_+} \right) \right]f^-_{E,\kappa}(x) = 0  \\
& \left( \frac{1}{2}\frac{d}{dx}-\frac{\kappa}{x} \right)f^-_{E,\kappa}(x) + x(\epsilon-\mu)f^+_{E,\kappa}(x) \\
&  + \frac{1}{x}\left[\epsilon - \mu\left(1-\frac{M_{ADM}}{r_+} \right)\right]f^+_{E,\kappa}(x) = 0
\end{split}
\end{equation}
We observe that if $M\rightarrow 0$ then the ratio $\frac{M_{ADM}}{r_+}\rightarrow\frac{M_s}{r_s}= \frac{1}{2}$ and $\mu\left(1-\frac{M_{ADM}}{r_+} \right)\rightarrow\frac{1}{2}\mu$ recovering the Schwarzschild case discussed in Ref. \cite{cota}.

In order to find the scattering modes, for the case $\epsilon>\mu$, it proves useful to introduce new radial wave functions $\hat f^\pm$, that consist in the following combination of the old wave functions $f^\pm$
\begin{equation}\label{sol8}
\begin{split}
& \hat f^+_{E,\kappa} = \frac{i}{2}\frac{f^+_{E,\kappa}}{\sqrt{\epsilon+\mu}} +\frac{1}{2}\frac{f^-_{E,\kappa}}{\sqrt{\epsilon-\mu}} \\
& \hat f^-_{E,\kappa} = -\frac{i}{2}\frac{f^+_{E,\kappa}}{\sqrt{\epsilon+\mu}} +\frac{1}{2}\frac{f^-_{E,\kappa}}{\sqrt{\epsilon-\mu}}
\end{split}
\end{equation}
Using the relation above and after some computations one arrives at the following equations satisfied by the functions $\hat f^+_{E,\kappa}$ and $\hat f^-_{E,\kappa}$, namely
\begin{equation}\label{sol9}
\openup 2\jot
\begin{split}
& \left[ \nu x\frac{d}{dx} +2\,i\left(\mu^2\left(1-\frac{M_{ADM}}{r_+} \right)-\epsilon^2-\nu^2x^2 \right) \right]\hat f^+_{E,\kappa} \\
& - \left( 2\kappa\nu-i\epsilon\mu\frac{M_{ADM}}{r_+} \right) \hat f^-_{E,\kappa} =0\\
& \left[ \nu x\frac{d}{dx} -2\,i\left(\mu^2\left(1-\frac{M_{ADM}}{r_+} \right)-\epsilon^2-\nu^2x^2 \right) \right]\hat f^-_{E,\kappa} \\
& -\left( 2\kappa\nu+i\epsilon\mu\frac{M_{ADM}}{r_+} \right) \hat f^+_{E,\kappa} =0
\end{split}
\end{equation}
where $\nu=\sqrt{\epsilon^2-\mu^2}$. Eq. (\ref{sol9}) can now be solved using Maple or Mathematica and the analytical solutions can be written as a combination of Whittaker functions
\begin{equation}\label{sol10}
\begin{split}
& \hat f^+_{E,\kappa} = C_1\frac{1}{x}M_{\rho_+,s}(z)+ C_2\frac{1}{x}W_{\rho_+,s}(z)\\
& \hat f^-_{E,\kappa} = \frac{1}{\kappa^2+\lambda^2}\left[ (s-i\alpha)(\kappa+i\lambda)C_1\frac{1}{x}M_{\rho_-,s}(z) \right. \\
& \qquad\qquad \qquad\qquad \left.-(\kappa+i\lambda)C_2\frac{1}{x}W_{\rho_-,s}(z)\right]
\end{split}
\end{equation}
where $z=2i\nu x^2$ and the following parameters were introduced
\begin{equation}\label{sol11}
\begin{split}
& s=\left[\kappa^2+\mu^2\left(1-\frac{M_{ADM}}{r_+} \right)^2 -\epsilon^2 \right]^{\frac{1}{2}}, \quad \lambda=\frac{\epsilon\mu}{\nu}\cdot\frac{M_{ADM}}{r_+} \\
& \rho_{\pm}=\mp\frac{1}{2} -i\alpha ,\qquad \alpha=\frac{1}{\nu}\left[ \epsilon^2 -\mu^2\left(1-\frac{M_{ADM}}{r_+} \right)\right]
\end{split}
\end{equation}

\subsubsection{Pure Bardeen black hole case}

As already mentioned, by choosing $M_s=0$ and $\gamma=3$ in eq. (\ref{ds2}) the original Bardeen black hole solution is recovered. This solution is a black hole with two distinct horizons only if $|Q|<4/\sqrt{27}$, it has degenerate horisons if $Q=4/\sqrt{27}$ and for $Q>4/\sqrt{27}$ there are no horizons present. Because the function $h(r)$ has now a more simpler form,
\begin{equation}\label{ds2b}
h(r)=1-\frac{2Mr^2}{\sqrt{(r^2+Q^2)^3}}
\end{equation}
one can find and write an analytical expression for the location of the black hole outer horizon, denoted from now on by $r_+$. Moreover, one can easily show that $r_+=M\,f(Q/M)$, where the function $f$ depends only on the ratio $Q/M$. The existence of horizons implies the constraint $|Q|/M\leq 4/\sqrt{27}$ \cite{borde}.

Following the same steps as in section \ref{sec.scatt}, one finds the same scattering modes as given by eq. (\ref{sol10}), but with the new parameters
\begin{equation}\label{sol11a}
\begin{split}
& s=\left[\kappa^2+\mu^2\left(1-\frac{M}{r_+} \right)^2 -\epsilon^2 \right]^{\frac{1}{2}}, \quad \lambda=\frac{\epsilon\mu}{\nu}\cdot\frac{M}{r_+} \\
& \rho_{\pm}=\mp\frac{1}{2} -i\alpha ,\qquad \alpha=\frac{1}{\nu}\left[ \epsilon^2 -\mu^2\left(1-\frac{M}{r_+} \right)\right]
\end{split}
\end{equation}
We will see in section \ref{sec.results} that this is enough to produce a noticeable difference in the scattering patterns.

\subsection{Analytical phase shifts and scattering cross sections}

In spinor wave scattering theory \cite{rose,Landau} the differential scattering cross section for an unpolarized incident beam is the sum of the squares of two scalar functions
\begin{equation}
\frac{d\sigma}{d\Omega}=|f(\theta)|^2 + |g(\theta)|^2
\end{equation}
that depend only on the scattering angle $\theta$:
\begin{equation}\label{sum1}
\begin{split}
&f(\theta)=\sum_{l=0}^{\infty}\frac{1}{2ip}\left[ (l+1)(e^{2i\delta_{-l-1}}-1)+ l(e^{2i\delta_l}-1)\right]\,P_l^0(\cos \theta) \\
&g(\theta)=\sum_{l=1}^{\infty}\frac{1}{2ip}\left[e^{2i\delta_{-l-1}}-e^{2i\delta_l}  \right]\,P_l^1(\cos\theta)
\end{split}
\end{equation}
were $p$ is the incident momentum and $P_l^0(\cos\theta),\,P_l^1(\cos\theta) $ stand for Legendre and associated Legendre polynomials, which are special cases of Legendre functions \cite{NIST}.

The phase shifts $\delta_l$ can be computed as in Refs. \cite{sporea1,sporea2} by applying the partial wave method on the scattering modes (\ref{sol10}). The asymptotic form of the radial wave functions $f^\pm$ can be written as \cite{sporea1}
\begin{equation}\label{arg}
\left(
\begin{array}{c}f^+_{E,\kappa}\\
f^-_{E,\kappa}
\end{array}\right)
\propto \begin{array}{c}
\sqrt{E+m}\,\sin\\
\sqrt{E-m}\,\cos
\end{array}\left( pr-\frac{\pi l}{2} +\delta_{\kappa}+\vartheta(r)\right)
\end{equation}
were $\vartheta(r)=-p r_++\alpha \ln [2p(r-r_+)]$ represents a radially dependent phase that is independent of any angular quantum numbers and thus dose not contribute to the scattering cross sections and may be neglected as done in the Dirac-Coulomb case \cite{dolan, Landau}.

The resultant final form for the point-independent phase shifts $\delta_\kappa$ is given by the following expression (see also Appendix A)
\begin{equation}\label{pshift}
e^{2i\delta_{\kappa}}=\frac{\kappa-i\lambda}{s-i\alpha}\cdot\frac{\Gamma(1+s-i\alpha)}{\Gamma(1+s+i\alpha)} e^{i\pi(l-s)}
\end{equation}
where we used for $\kappa$ the same sign convention as in \cite{Landau} such that $\kappa=\pm(j+1/2)$ and $l=|\kappa|-(1-sign\,\kappa)/2$.

The series (\ref{sum1}) are poorly convergent as a direct consequence of the singularity present at $\theta=0$ that requires an infinite number of Legendre polynomials to describe it. In order to make the series more convergent one can define the $m$th reduced series \begin{equation}\label{sch.ex1}
\openup 2\jot
\begin{split}
& (1-\cos\theta)^{m_1}f(\theta)= \sum_{l\ge 0} a_l^{(m_1)} P_l(cos\theta) \\
&(1-\cos\theta)^{m_2}g(\theta)= \sum_{l\ge 1} b_l^{(m_2)} P_l^1(\cos\theta)
\end{split}
\end{equation}
as first proposed in \cite{Yennie} and more recently used in \cite{dolan, sporea1,sporea2}. The new coefficients $a_l^{(i)}$ and $b_l^{(i)}$ are computed using the recurrence relations
\begin{equation}\label{sch.ex3}
\openup 2\jot
\begin{split}
&a_l^{(i+1)}=a_l^{(i)}-\frac{l+1}{2l+3}a_{l+1}^{(i)}-\frac{l}{2l-1}a_{l-1}^{(i)} \\
&b_l^{(i+1)}=b_l^{(i)}-\frac{l+2}{2l+3}b_{l+1}^{(i)}-\frac{l-1}{2l-1}b_{l-1}^{(i)}
\end{split}
\end{equation}
with $a_l^{(0)}=\left[ (l+1)(e^{2i\delta_{-l-1}}-1)+ l(e^{2i\delta_{l}}-1)\right]/2ip$ and $b_l^{(0)}=\left[e^{2i\delta_{-l-1}}-e^{2i\delta_{l}}  \right]/2ip$ taken from eq. (\ref{sum1}). We found that using only two iterations $m_1=2$ for the function $f(\theta)$ and one iteration $m_2=1$ for $g(\theta)$ it is sufficient to make the series convergent enough without distorting too much the analytical results.

\section{Results and discussion}\label{sec.results}

In this section we present and discuss the main features of fermion scattering by Bardeen regular black holes and also by a Bardeen-class of black holes. The analysis will focus on scattering by small or micro black holes (with $M_{ADM}\sim 10^{15} - 10^{22}$ Kg) because in this case the glory and orbiting scattering phenomena are shown to be significant.

In labeling the figures we will make use of the following parameters: $v=p/E$ the speed of incident fermions; $EM$ that can be seen as a dimensionless measure of the gravitational coupling because (restoring the units) it forms the dimensionless quantity: $\varepsilon=\frac{GEM}{\hbar c^3}=\frac{\pi r_s}{\lambda v}$, with $r_s$ the Schwazschild radius and $\lambda=h/p$ the associated quantum particle wavelength; and the ratios $q=Q/M_s$ and $g=M/M_s$ that appear when writing the black hole horizon radius as $r_+=M_s\,f(\frac{M}{M_s},\,\frac{Q}{M_s})$. Moreover, sice in the asymptotic zone $E=\sqrt{m^2+p^2}$, one can easily show that $mM=EM\sqrt{1-v^2}$ such that the condition $EM\geq mM$ is always satisfied. In the following analysis we will take $\gamma=3$ in all the plots (excepting those in Fig. \ref{fig4a}), because the same conclusions are obtained also for the cases with $\gamma\neq3$.

\begin{figure*}
	\centering
	\includegraphics[scale=0.44]{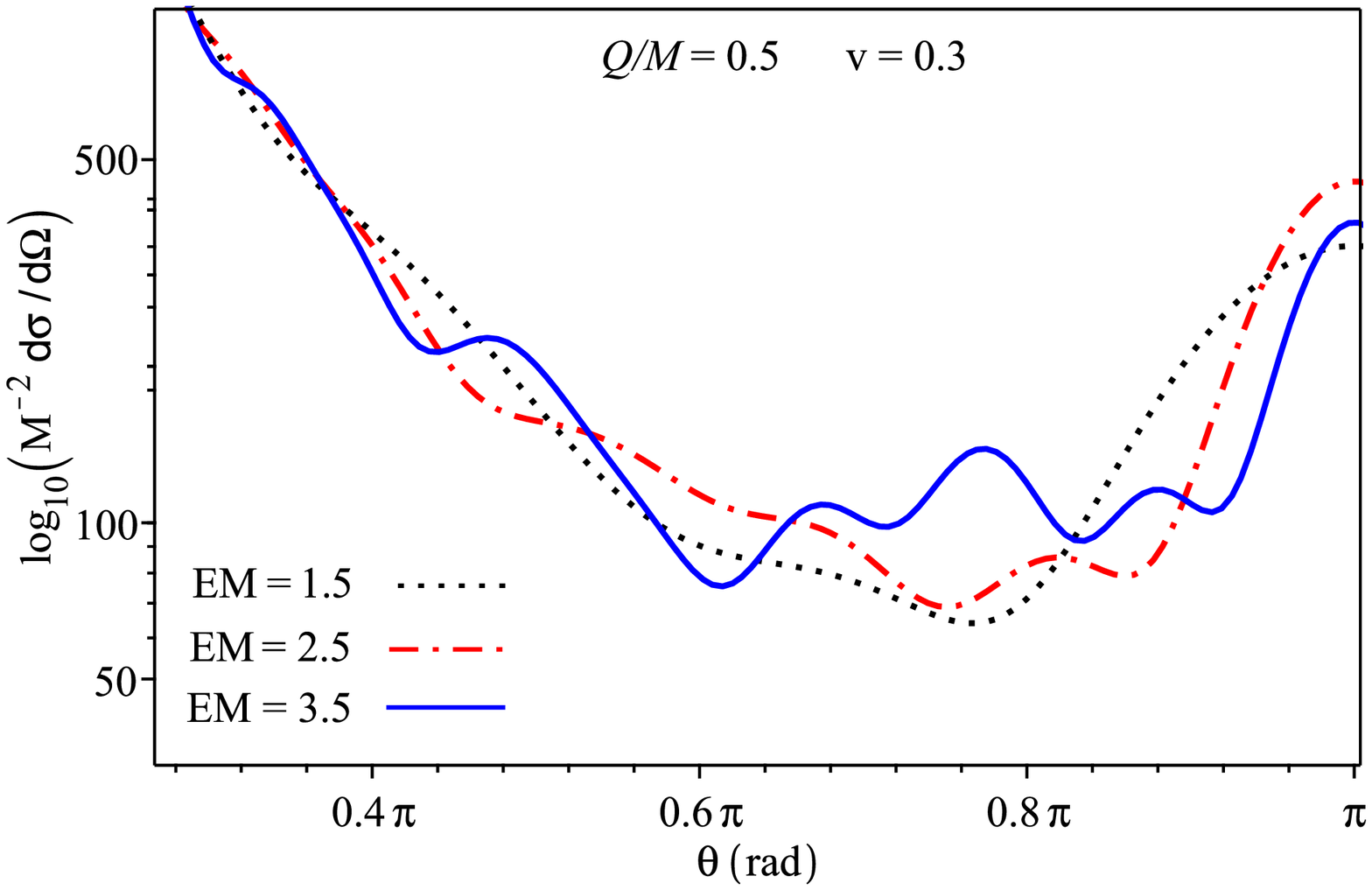}
	\includegraphics[scale=0.44]{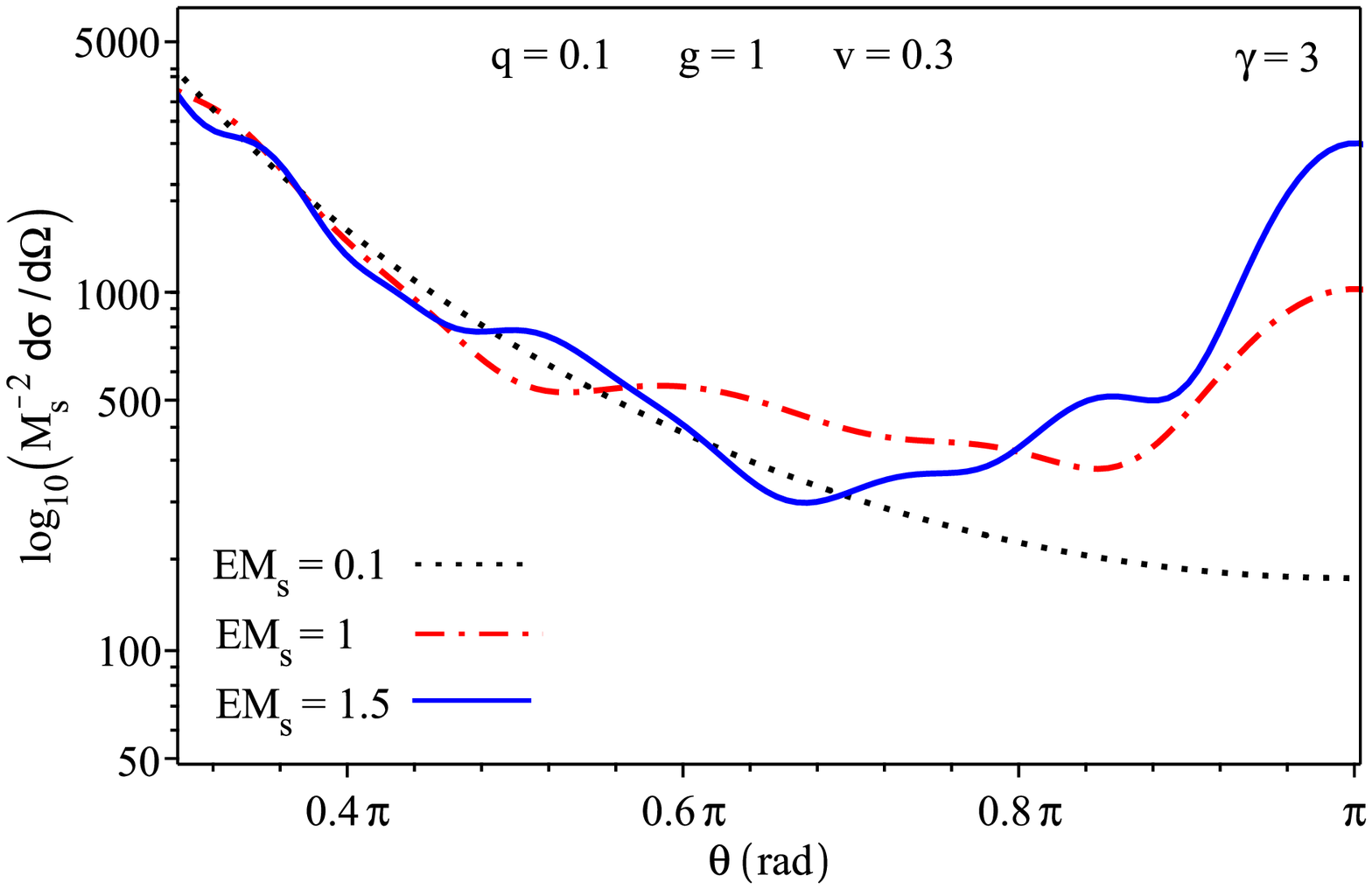}
	\includegraphics[scale=0.44]{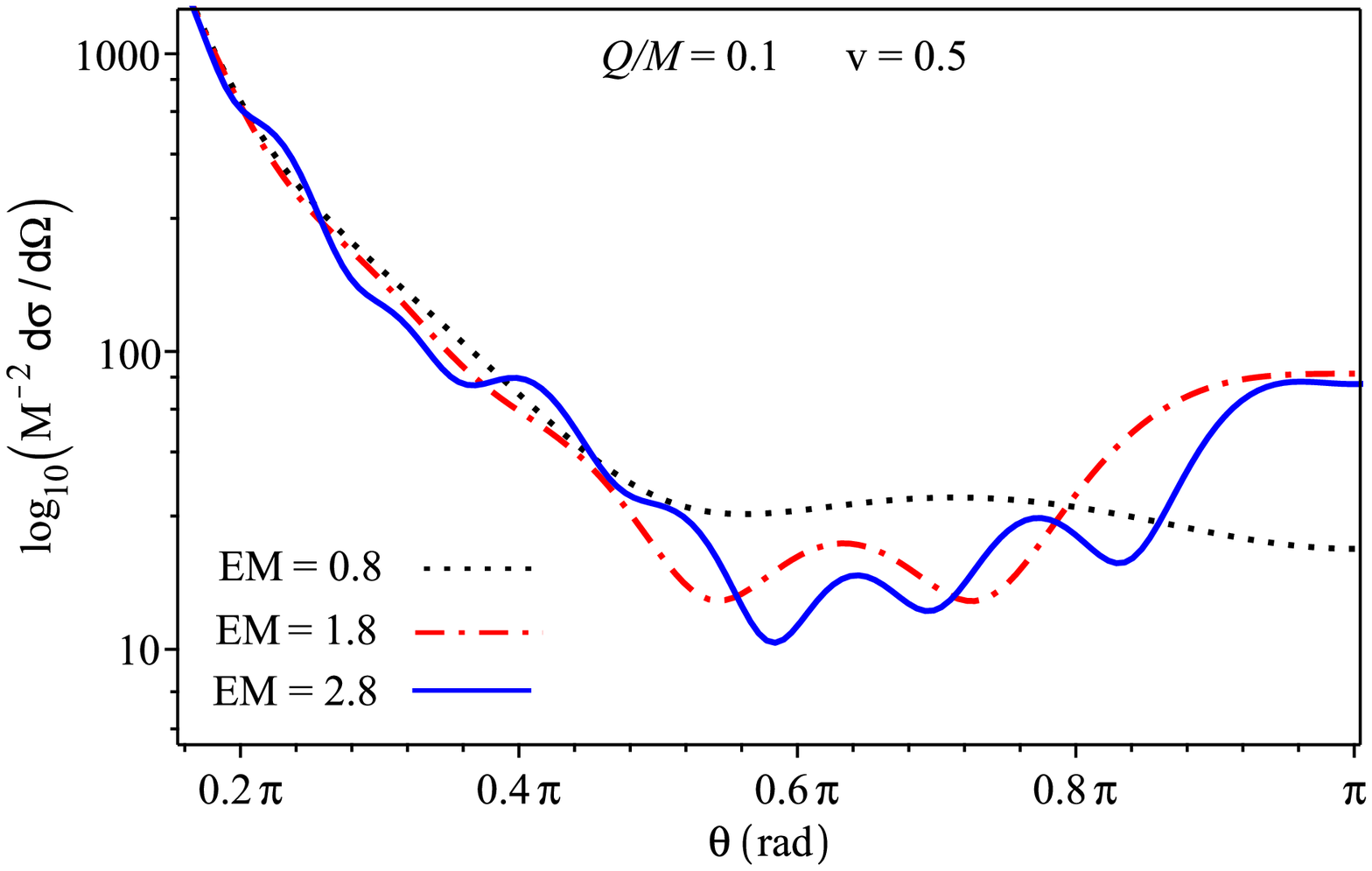}
	\includegraphics[scale=0.44]{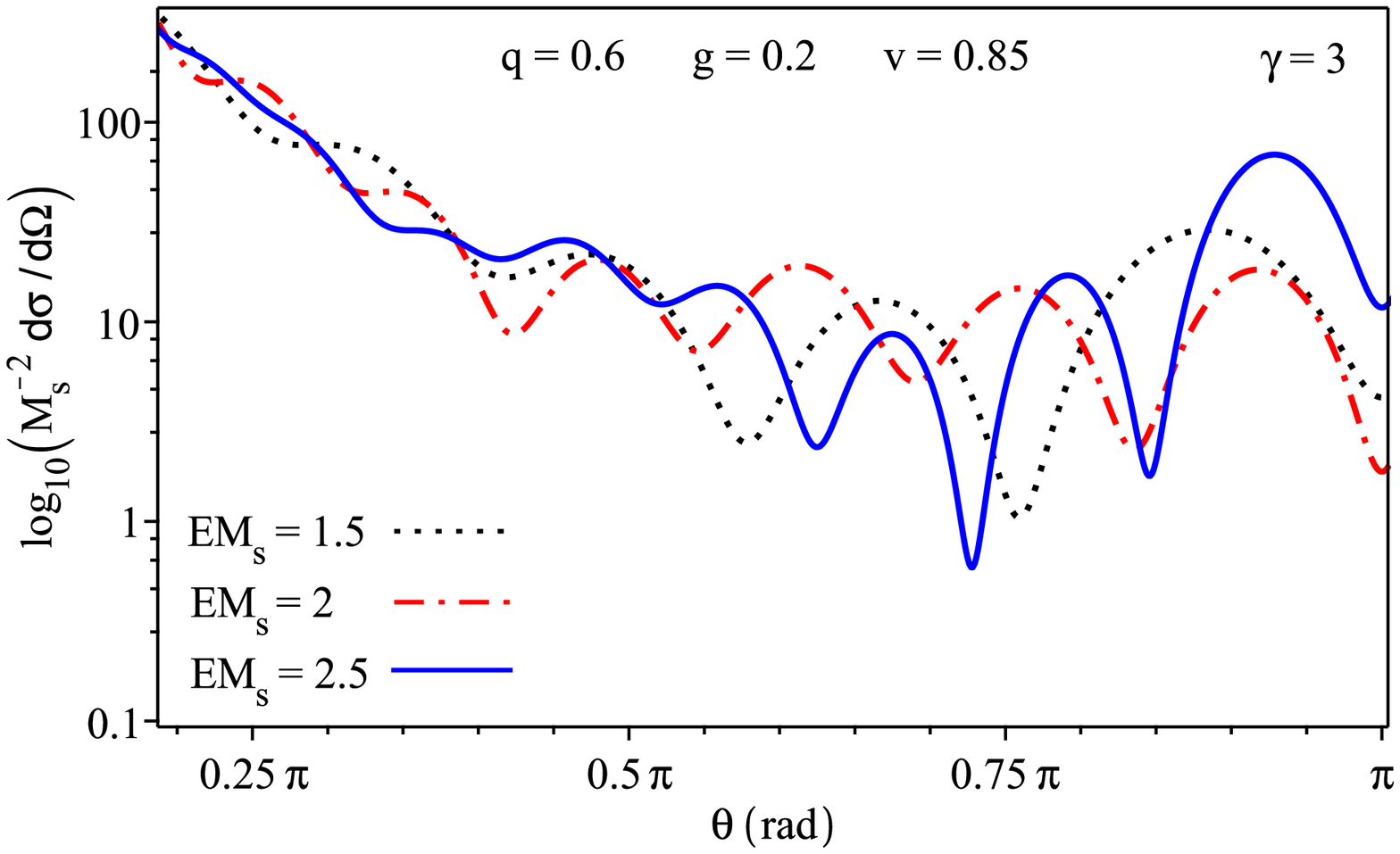}
	\includegraphics[scale=0.44]{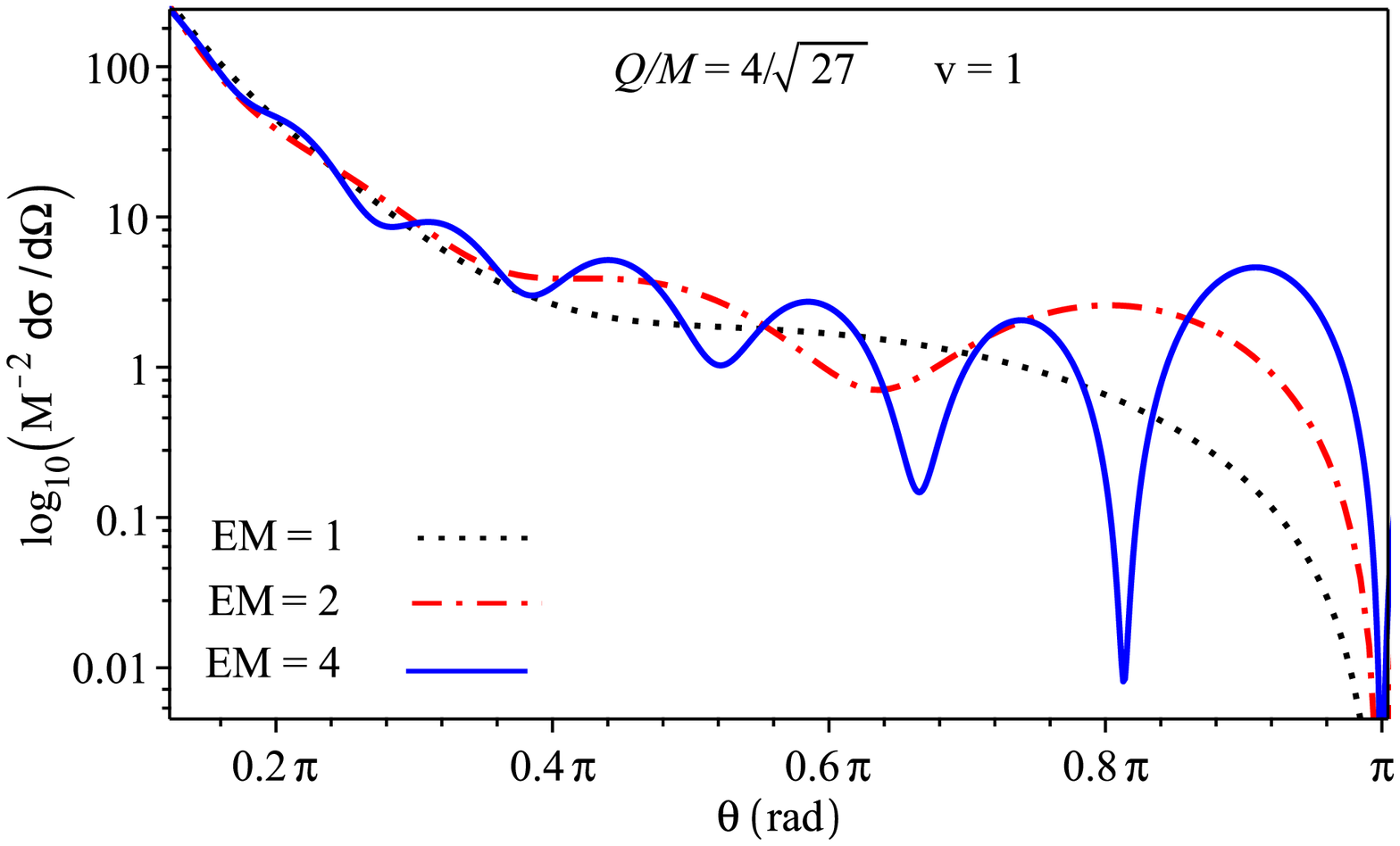}
	\includegraphics[scale=0.44]{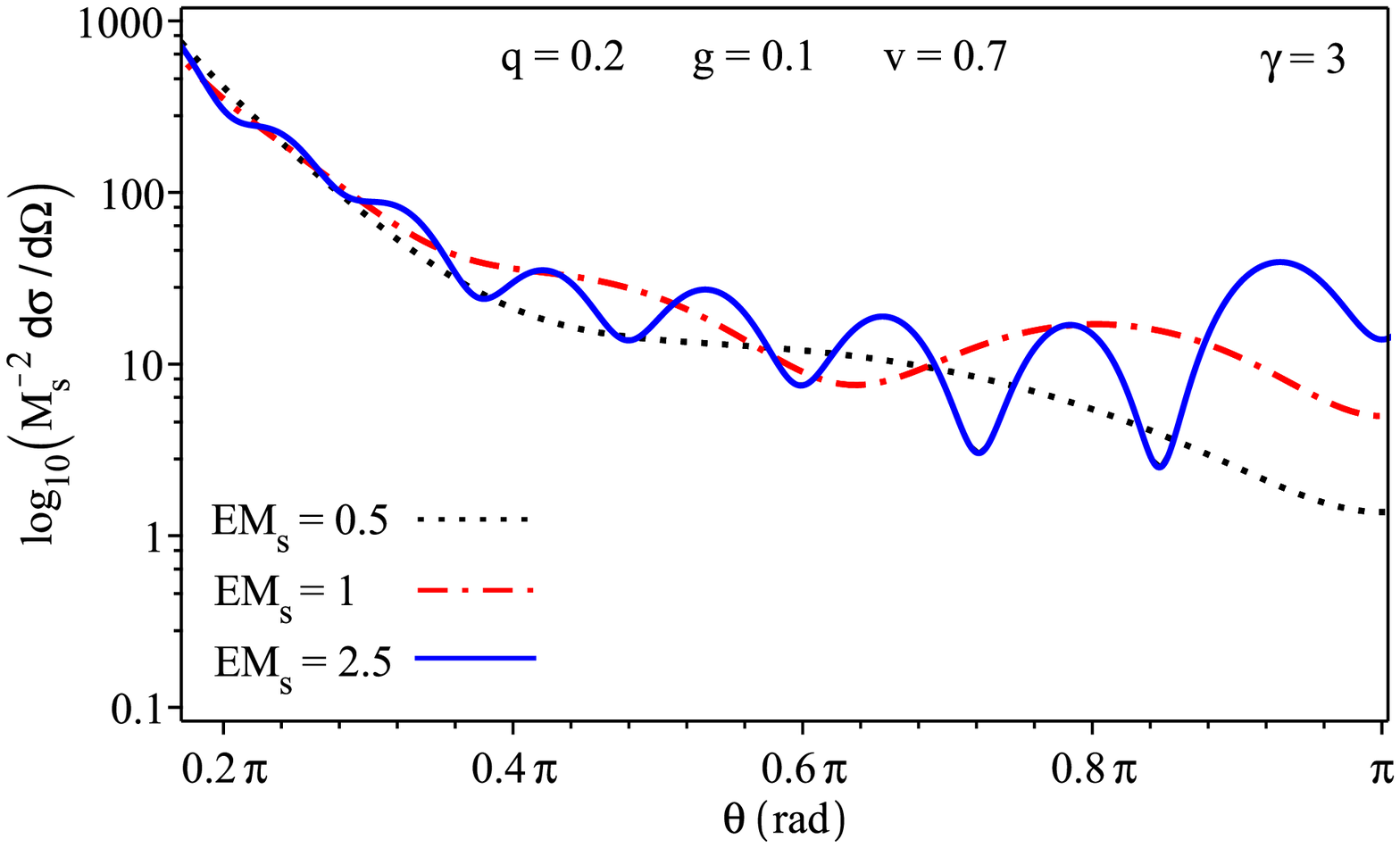}
	\caption{(color online). Plot of the differential scattering cross section as a function of the scattering angle for a regular Bardeen black hole, eq. \ref{ds2b} (left panels) and for a Bardeen-class black bole, eq. \ref{ds2} (right panels). The phenomena of glory (scattering in the backward direction) and spiral scattering (oscillations in the scattering intensity) are present for both types of black holes. The value $q/M_{em}=4/\sqrt{27}$ corresponds to a degenerate regular Bardeen black hole (it has only one horizon). The parameter $q$ gives the ratio $q=Q/M_s$ and $g=M/M_s$ is the ratio between the magnetic monopole mass and the "Schwarzschild mass". }
	\label{fig1}
\end{figure*}

In Fig. \ref{fig1} the differential scattering cross section as a function of the scattering angle is presented for fixed values of the ratios: $Q/M$ (left panels corresponding to pure Bardeen case); $q=Q/M_s$ and $g=M/M_s$; a fixed value of the speed ($v$) of incoming fermions, while the parameter $EM$ (or $EM_s$ for Bardenn-class) takes different values. As it can be seen in Fig. \ref{fig1} the scattering pattern takes a simple form for small values of $EM$. However, as the value of $EM$ is increased more complex scattering patterns start to occur, including the presence of a maximum in the backward direction ($\theta=\pi$) known also as "glory" scattering \cite{Ford1, Ford2} and the presence of oscillations in the scattering intensity that give rise to orbiting or "spiral scattering" \cite{Matzner1, Matzner2} (that may occur when the particle's "classical" orbit passes the scattering center multiple times). As the speed of the incoming fermion is increased, the peak in the $\pi$-direction starts to move to the left and the scattering intensity maximum that was occurring at $\theta=\pi$ for non-relativistic fermions is transforming into a minimum if the fermions are massless ($v=1$). As it can be best seen in the bottom panels of Fig. \ref{fig1} the magnitude of the spiral scattering oscillations and their angular frequency are increasing with the mass of the black hole.

\begin{figure*}
	\centering
	\includegraphics[scale=0.44]{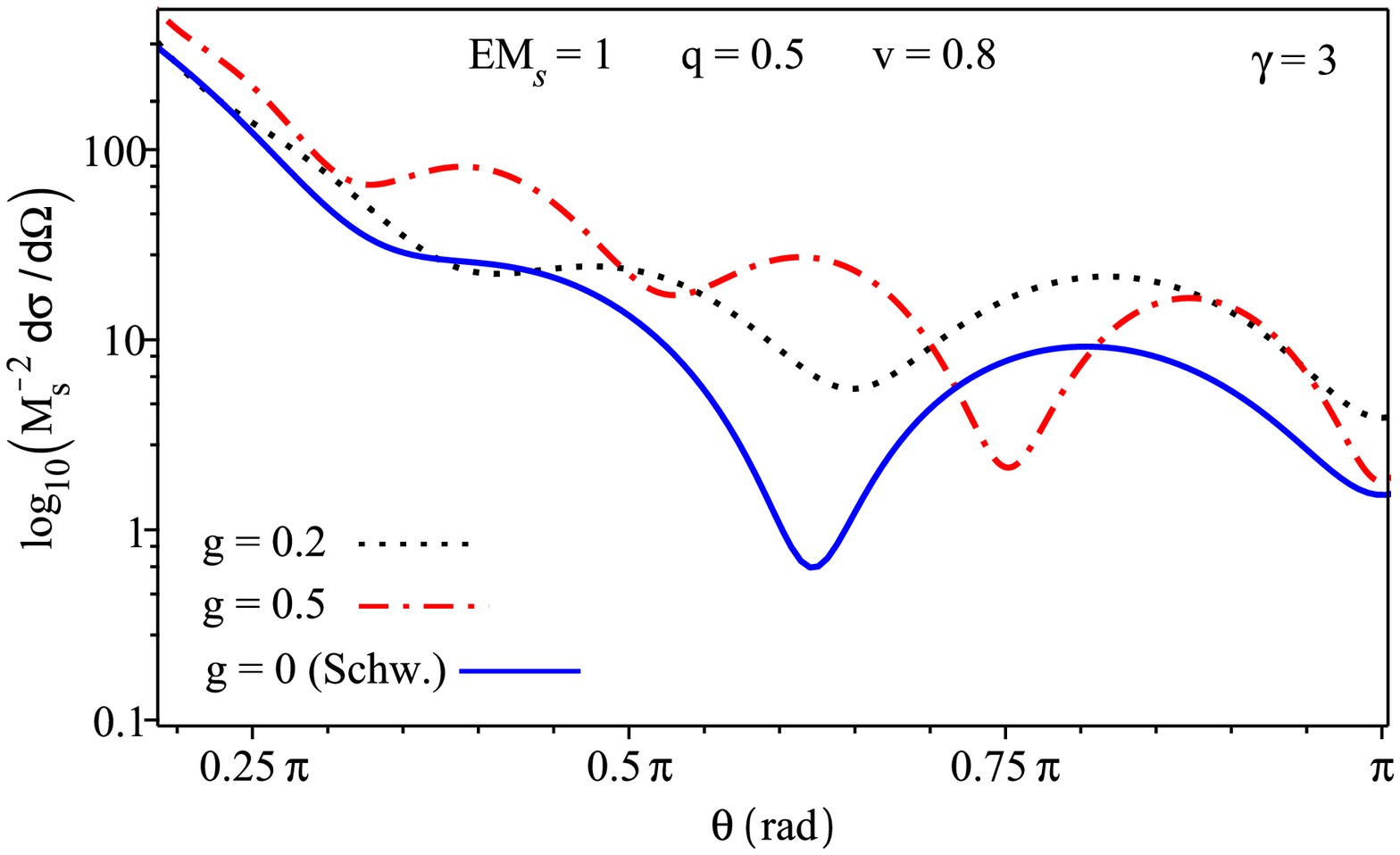}
	\includegraphics[scale=0.44]{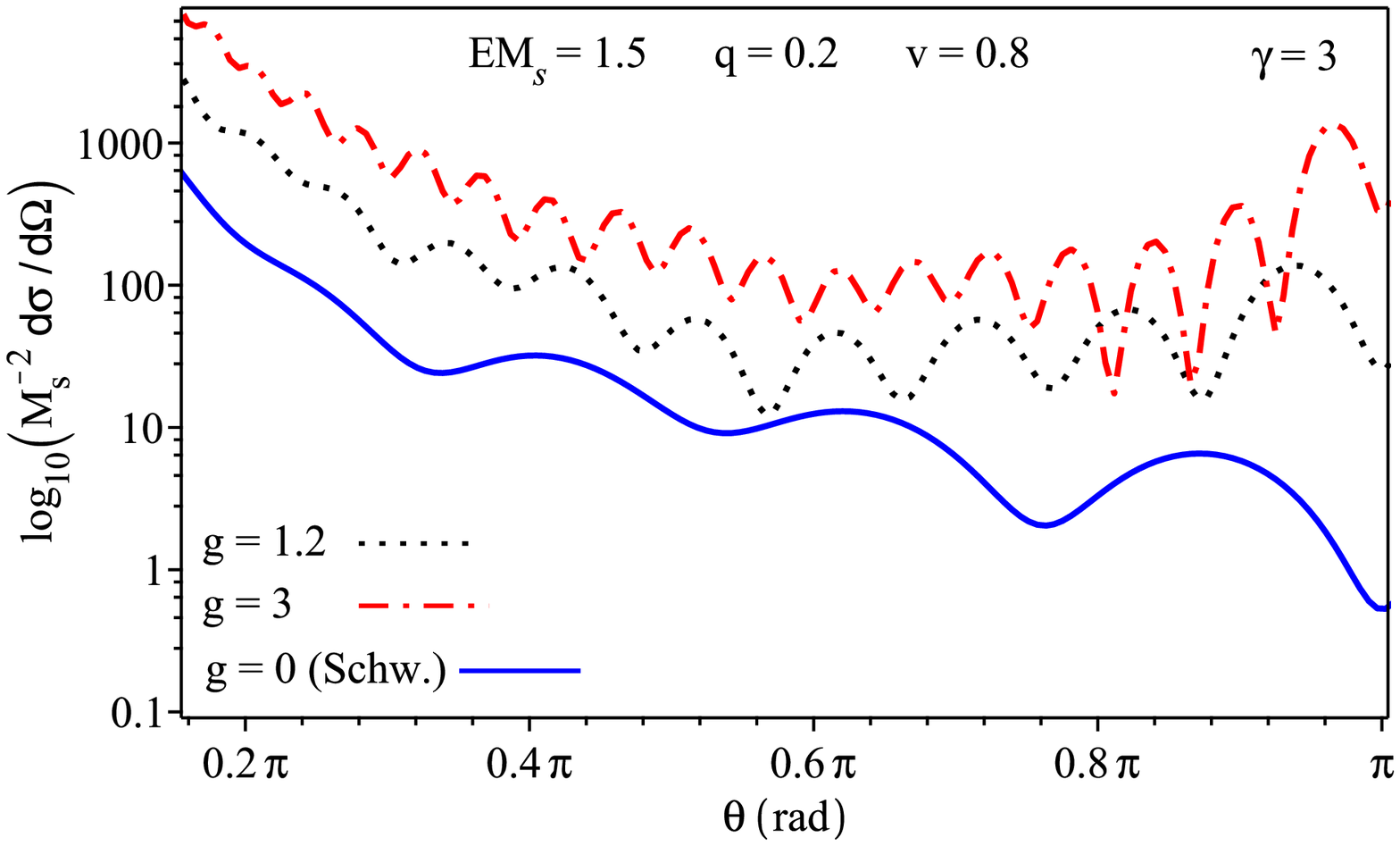}
	\includegraphics[scale=0.44]{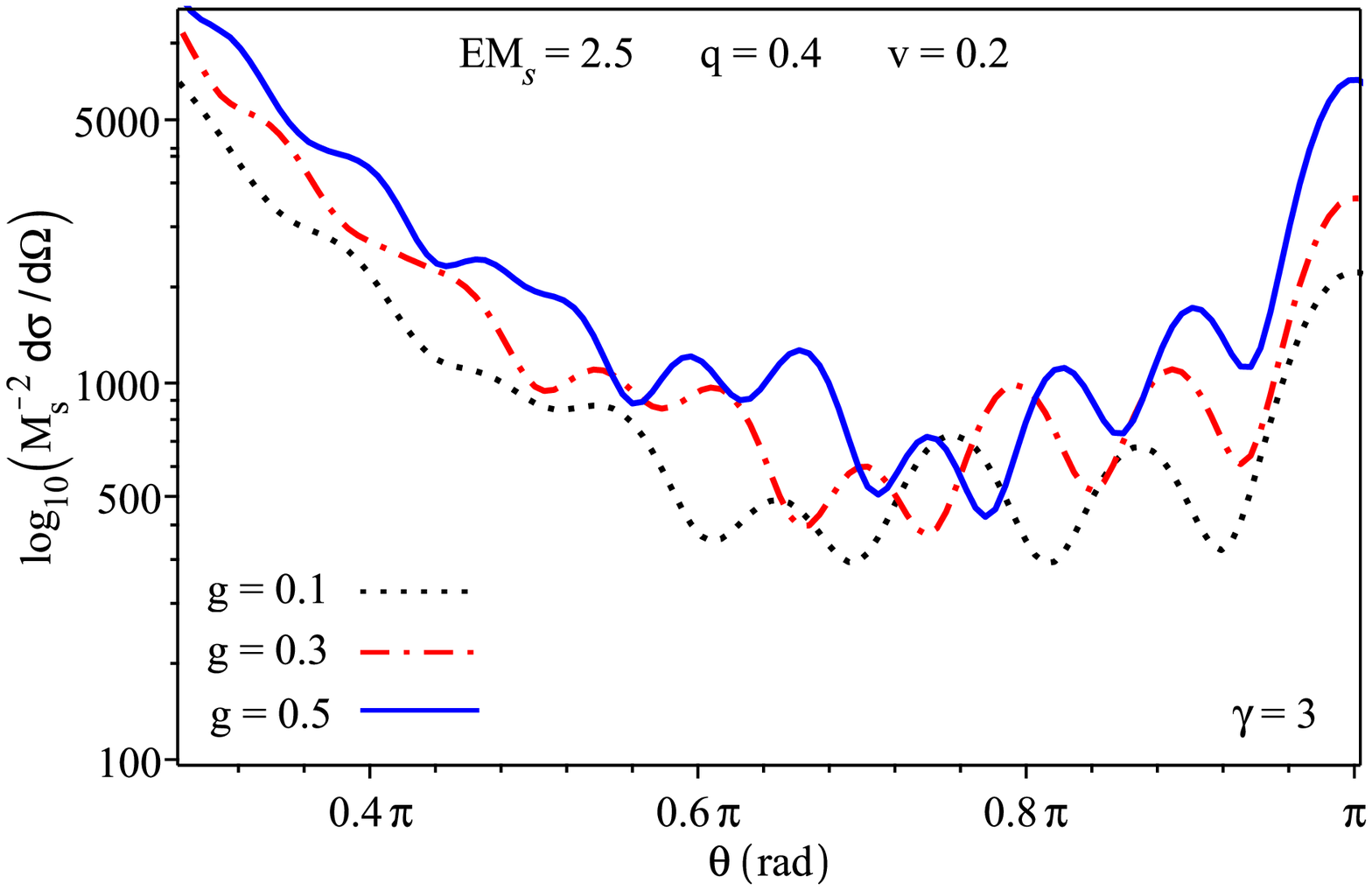}
	\includegraphics[scale=0.44]{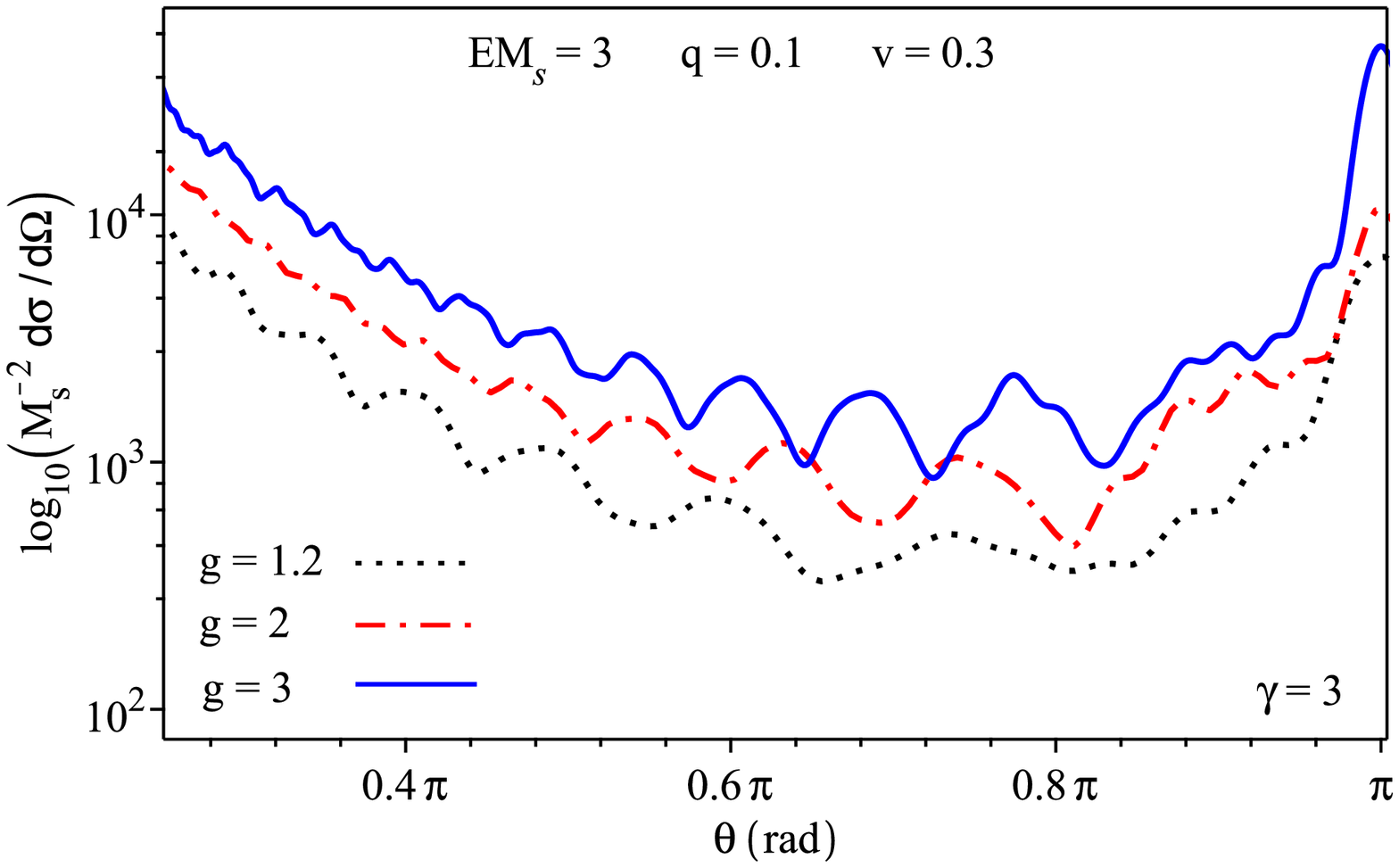}
	\caption{(color online). The variation of the scattering cross section with the parameter $g=M/M_s$ for given values of $EM_s$ and $v$. In the top panels comparison with the Schwarzschild scattering (for which $g=0$) is made. We can observe that as the value of $g$ is increasing the spiral scattering becomes more pronounced.}
	\label{fig3}
\end{figure*}

\begin{figure*}
	\centering
	\includegraphics[scale=0.44]{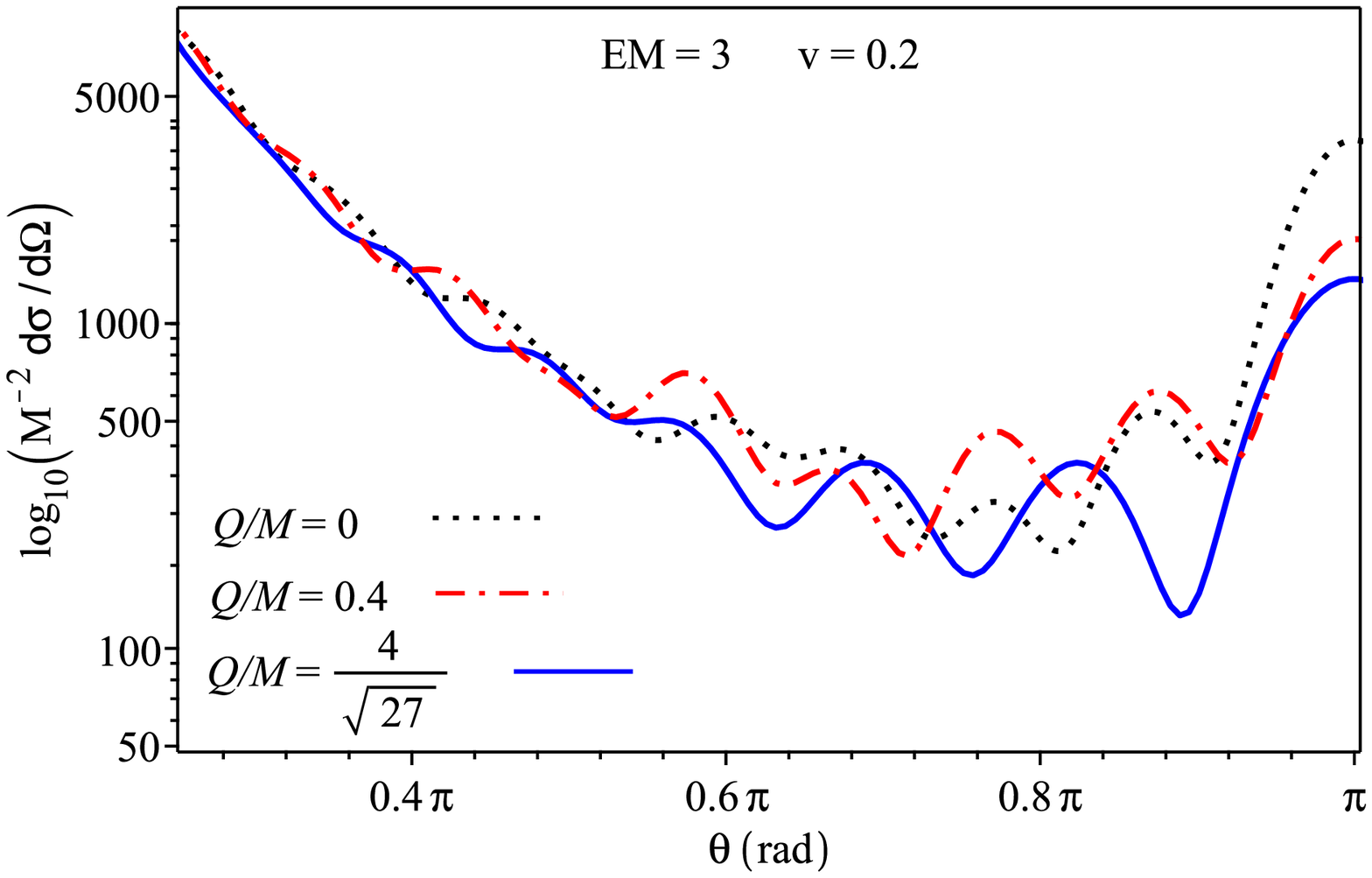}
	\includegraphics[scale=0.44]{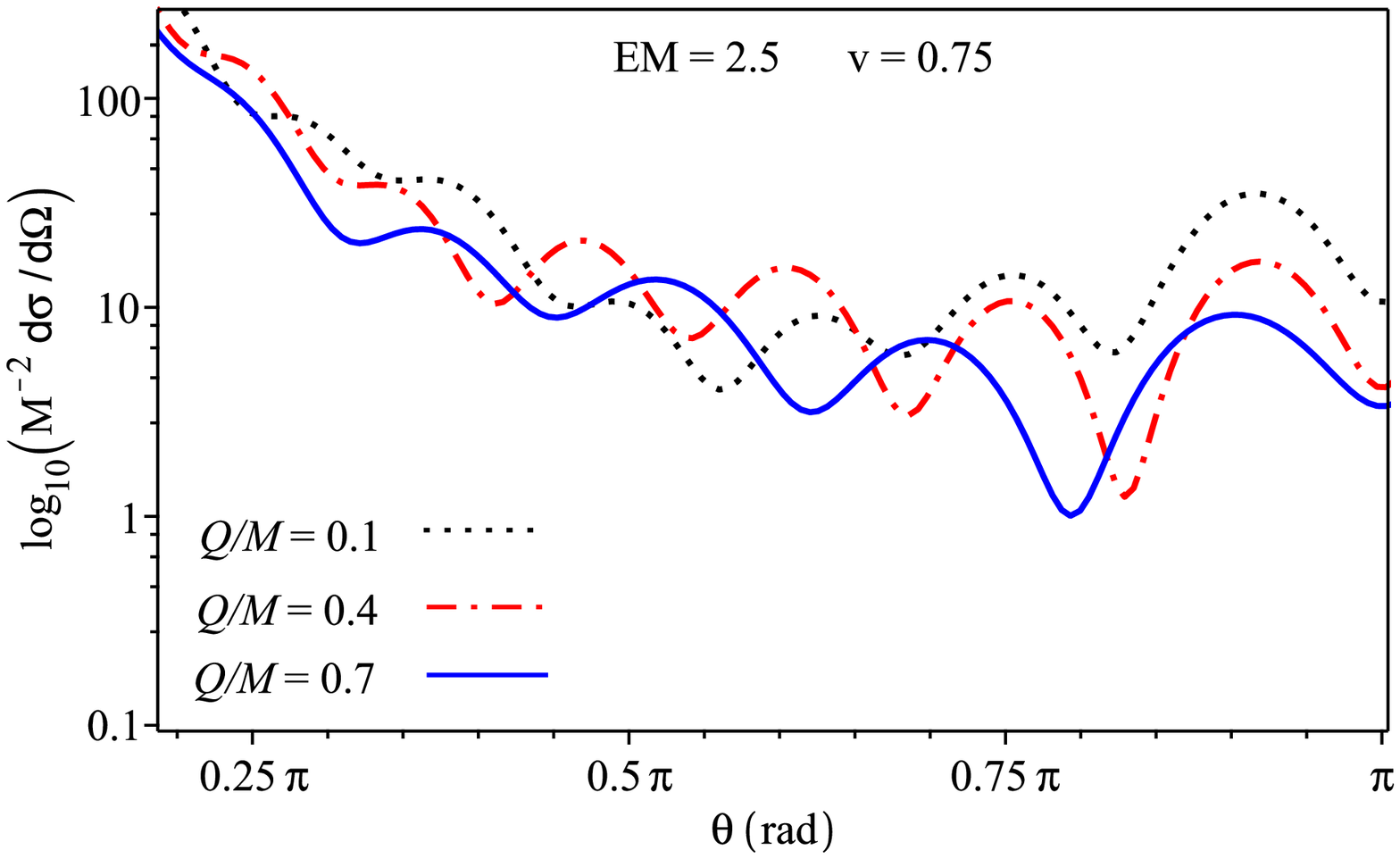}
	\caption{(color online). The scattering cross section for a regular Bardeen black hole for different values of $Q/M$, while the parameters $EM$ and $v$ are kept fixed. }
	\label{fig2}
\end{figure*}

\begin{figure*}
	\centering
	\includegraphics[scale=0.44]{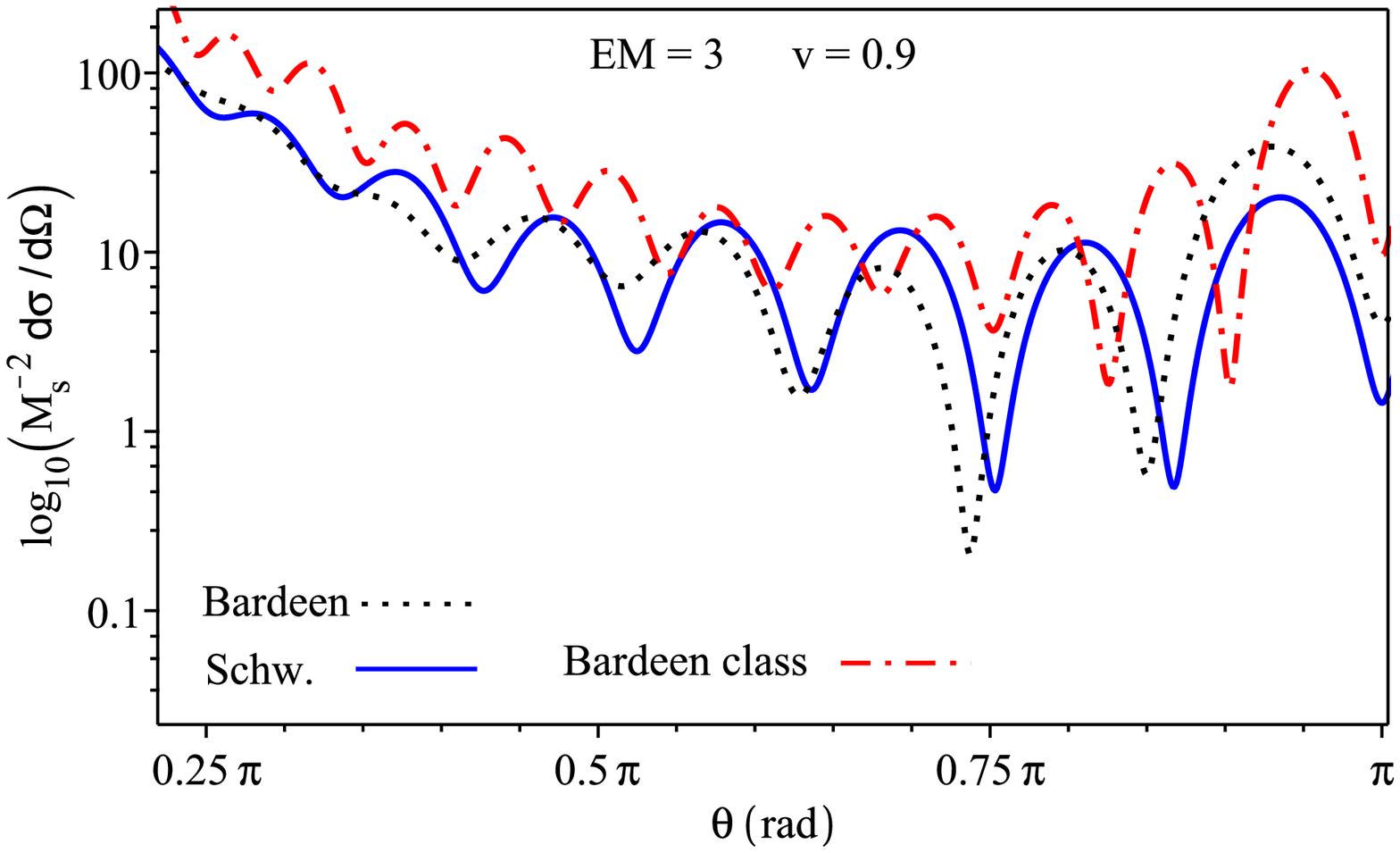}
	\includegraphics[scale=0.44]{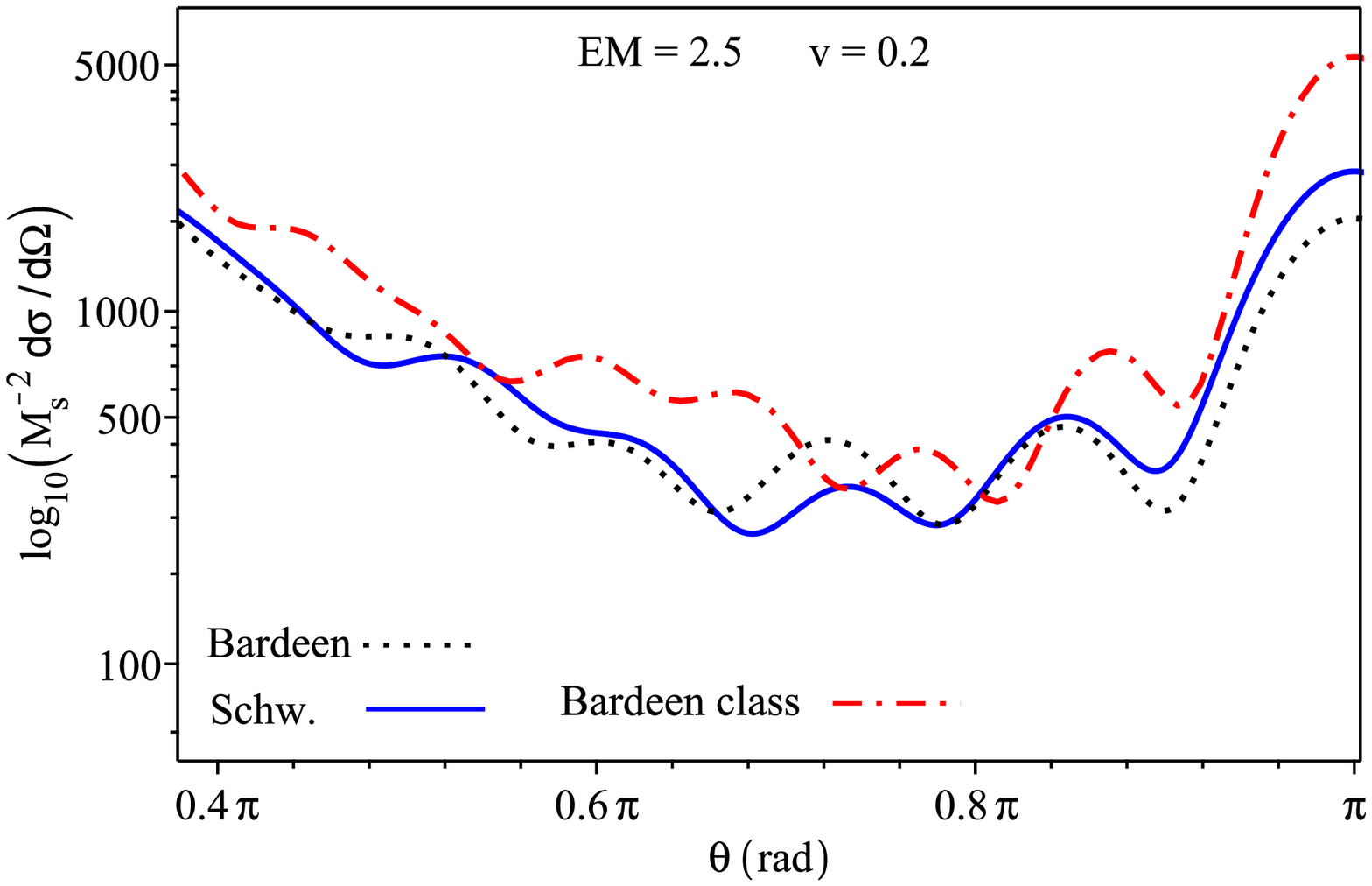}
	\caption{(color online). Comparison between the fermion scattering by a Schwarzschild black hole, a regular Bardeen black hole and a Bardeen-class black hole. For the {\it left panel} $Q/M=0.2$, $q=0.2$ and $g=0.5$ were used, while for the {\it right panel} $Q/M=0.4$, $q=0.1$ and $g=0.2$.  }
	\label{fig4}
\end{figure*}

The scattering cross section $d\sigma/d\Omega$ for the Bardeen-class black holes (\ref{ds2}) is plotted in Fig. \ref{fig3}. The left panels show the scattering patterns for Bardeen-class black holes that have the "Schwarzschild mass" bigger than the mass of the magnetic monopoles, while in the right side panels the opposite is true. From our analysis results that when compared with the Schwarzschild black hole, the scattering intensity in the backward direction ($\theta=\pi$) is higher for the scattering by a Bardeen-class black hole. From Fig. \ref{fig3} one can also observe that in all the plots as the value of $g$ is increasing more oscillations start to appear into the scattering intensity. Furthermore, if we assume $M_s$ fixed then it results that the $ADM$-mass of the black hole is increasing (which is equivalent to increasing the mass $M$, eq. \ref{ADM}, if $M_s$ remains constant) and as a consequence the angular frequency of the oscillations present in the scattering intensity are increasing as well. This feature was shown to be true also for fermion scattering by Schwarzschild and Reissner-Nordstr\"om black holes \cite{dolan,sporea1,sporea2,sporea3}. Thus one can wonder if the frequency of the oscillations of the spiral scattering are increasing with the black hole mass for any type of (spherically symmetric) black holes.

Comparing the scattering intensity, in Fig. \ref{fig2}, of a Bardeen regular black hole (blue and red curves) with that of a Schwarzschild black hole (the black dotted curve) we observe that the glory peak is higher for the scattering by a Schwarzschild black hole, while as the ratio $Q/M$ increases the maximum in the backward direction becomes lower and at the same time the frequency of oscillations in the spiral scattering are slightly decreasing  as well. The curve with $Q/M=4/\sqrt{27}$ corresponds to the Bardeen degenerate case, when only one horizon is present. When it will be possible to observe and measure the scattering of a beam of fermions by a black hole, then by analyzing the patterns in the scattering intensity, the type of black hole could be determined. If it is a Bardeen black hole, having a known mass, then the value of $Q$ can also be found.

In Fig. \ref{fig4} we plotted the differential scattering cross section $\frac{d\sigma}{d\Omega}$ as a function of the scattering angle $\theta$ for a Schwarzschild black hole (blue solid lines), a regular Bardeen black hole (black dotted lines) and a Bardeen-class black hole (red dash-dotted lines).

\begin{figure*}
	\centering
	\includegraphics[scale=0.44]{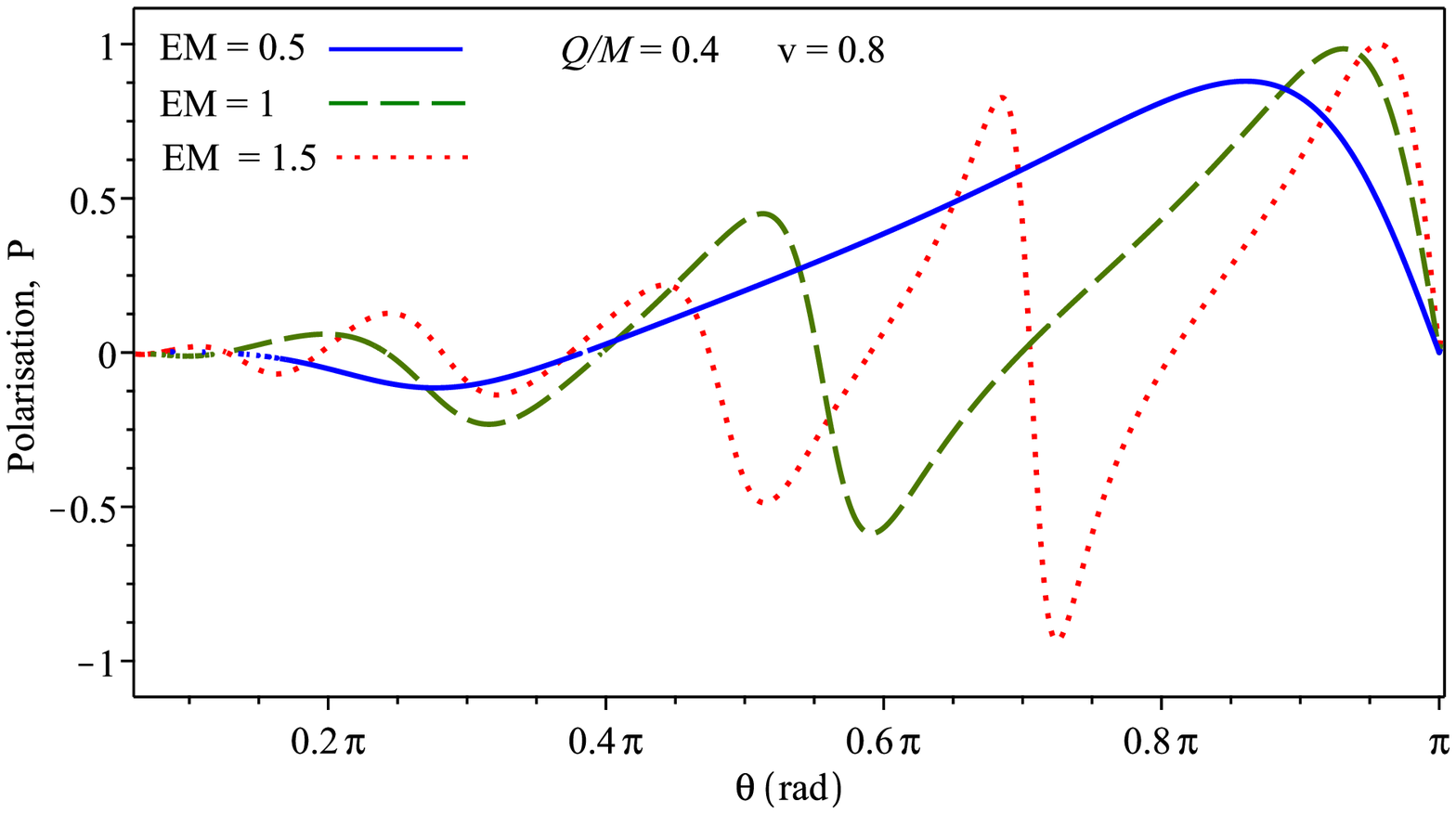}
	\includegraphics[scale=0.44]{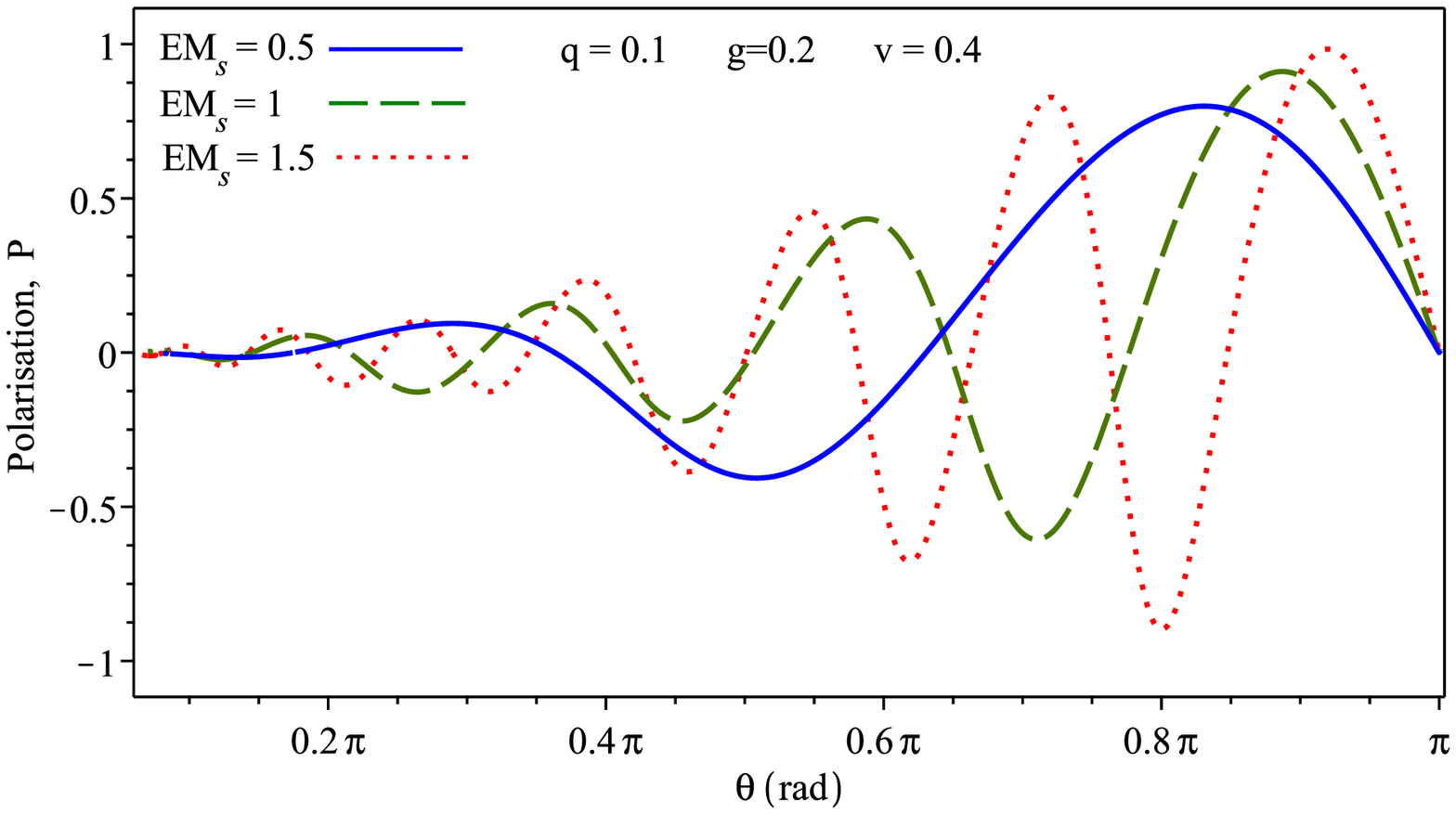}
	\includegraphics[scale=0.44]{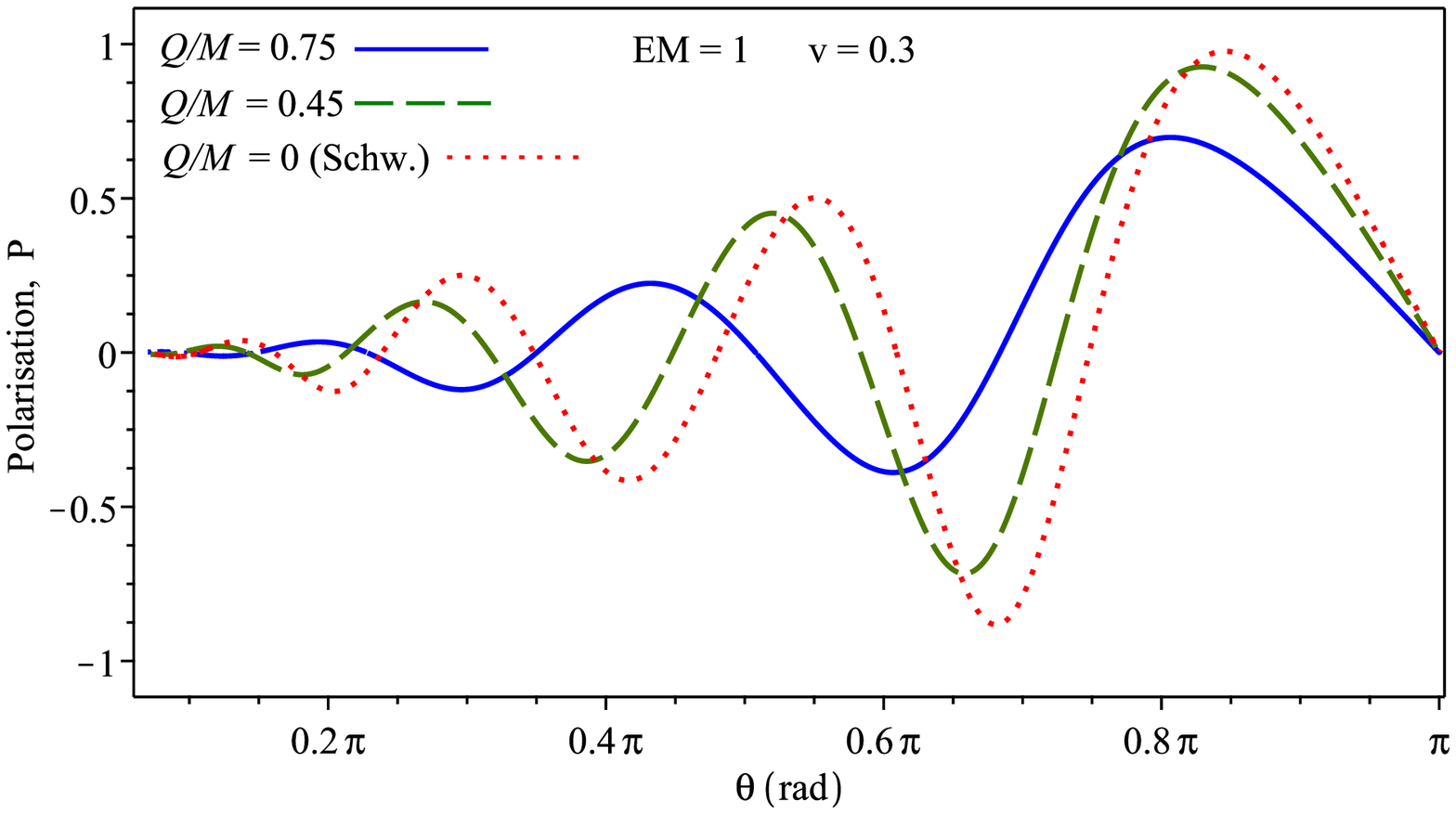}
	\includegraphics[scale=0.44]{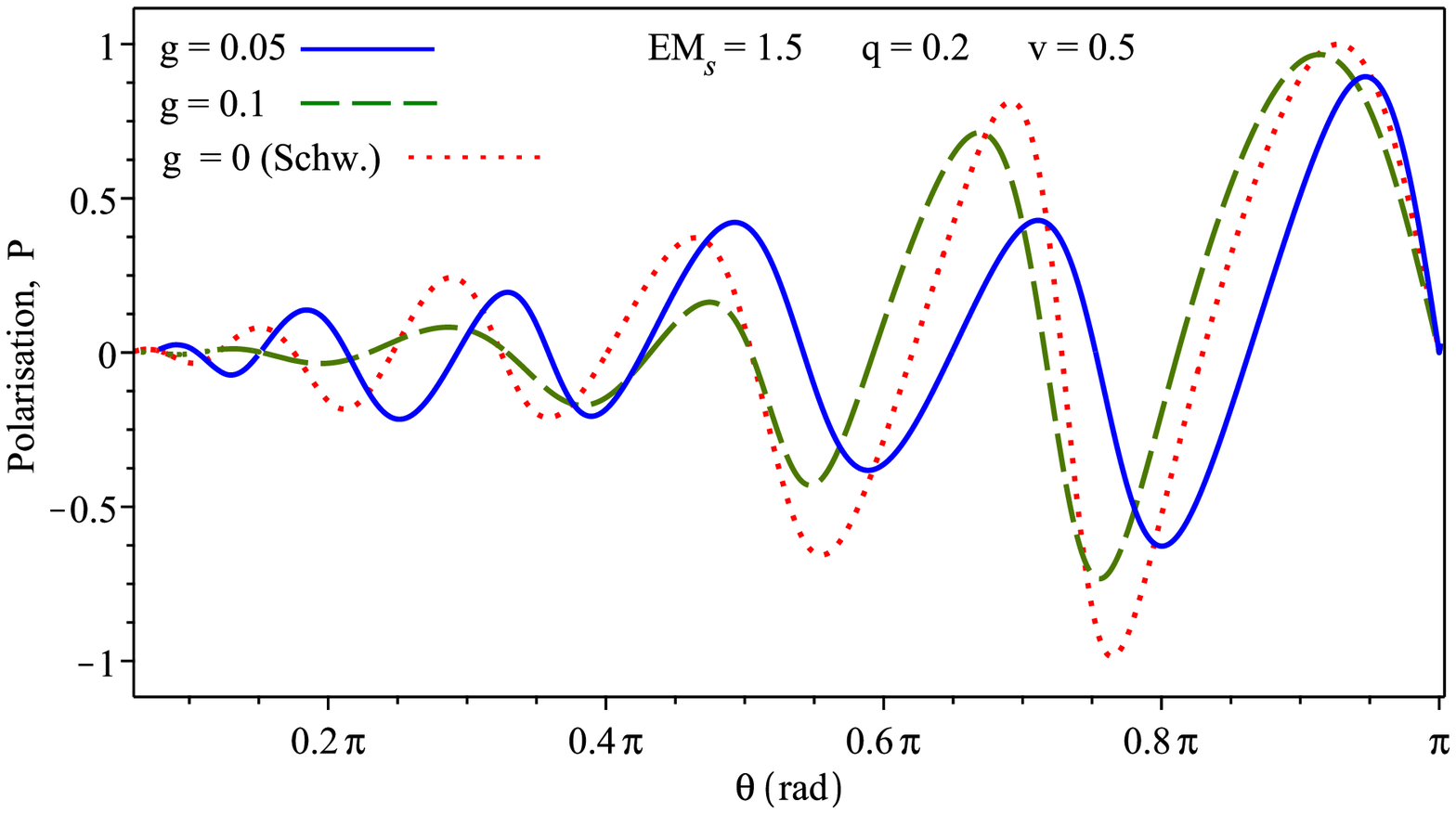}
	\caption{(color online). The partial polarization $\mathcal{P}$ as a function of the scattering angle $\theta$. {\it Top left panel:} typical Bardeen black hole (\ref{ds2b}) with $Q/M=0.4$ for incoming fermions with speed $v=0.8$ in units of $c$; {\it Top right panel:} Bardeen-class black hole (\ref{ds2}) with $g=M/M_s=0.2$ for incoming fermions with speed $v=0.4$ in units of $c$; {\it Bottom left panel:} comparing Schwarzschild and pure Bardeen black hole polarizations for fixed $EM=1$ and $v=0.3$; {\it Bottom right panel:} comparing Schwarzschild and Bardeen-class black hole polarizations for fixed $EM_s=1.5$, $q=Q/M_s=0.2$ and $v=0.5$.}
	\label{fig5}
\end{figure*}

\begin{figure*}
	\centering
	\includegraphics[scale=0.36]{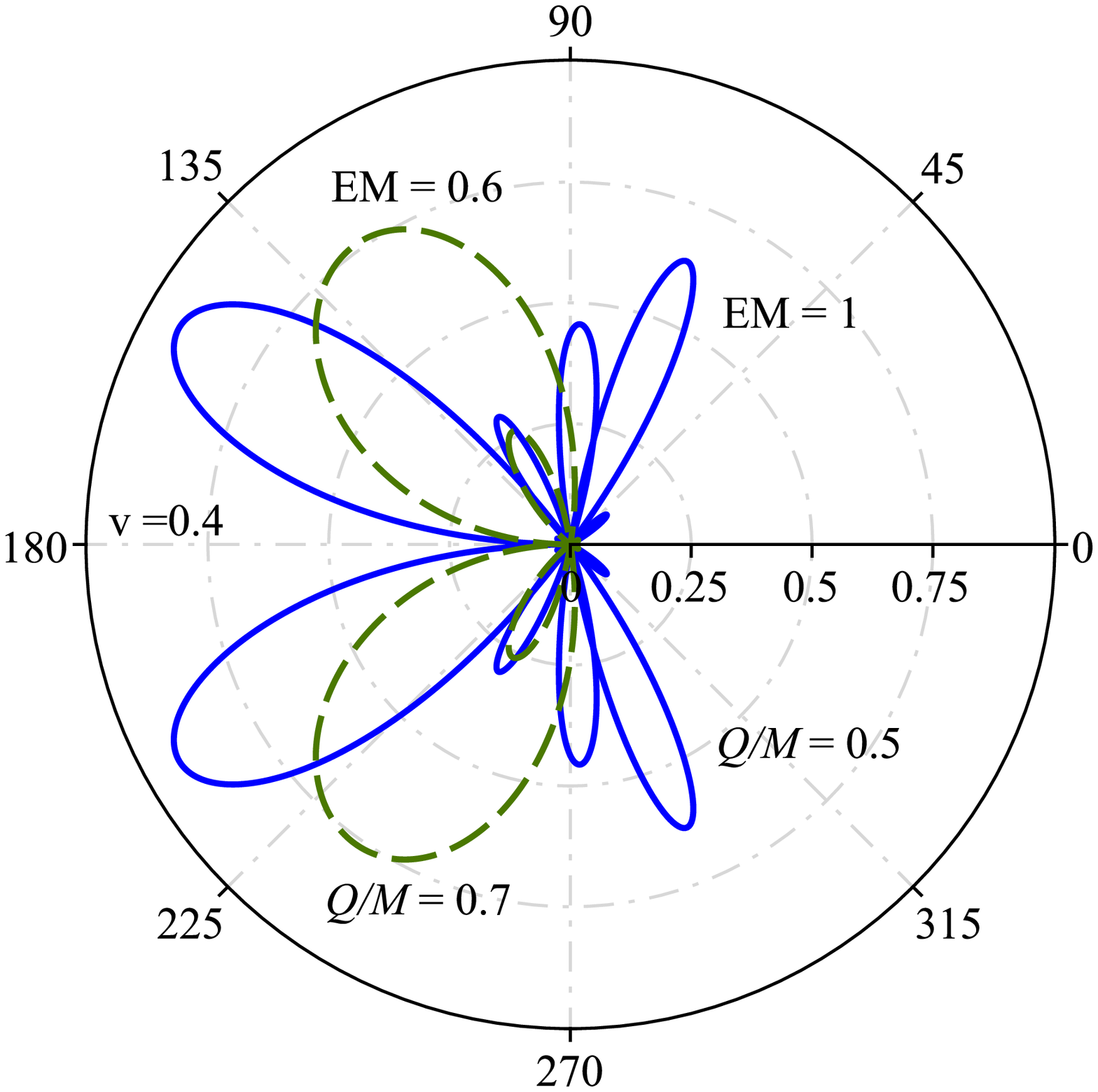}
	\qquad\qquad
	\includegraphics[scale=0.36]{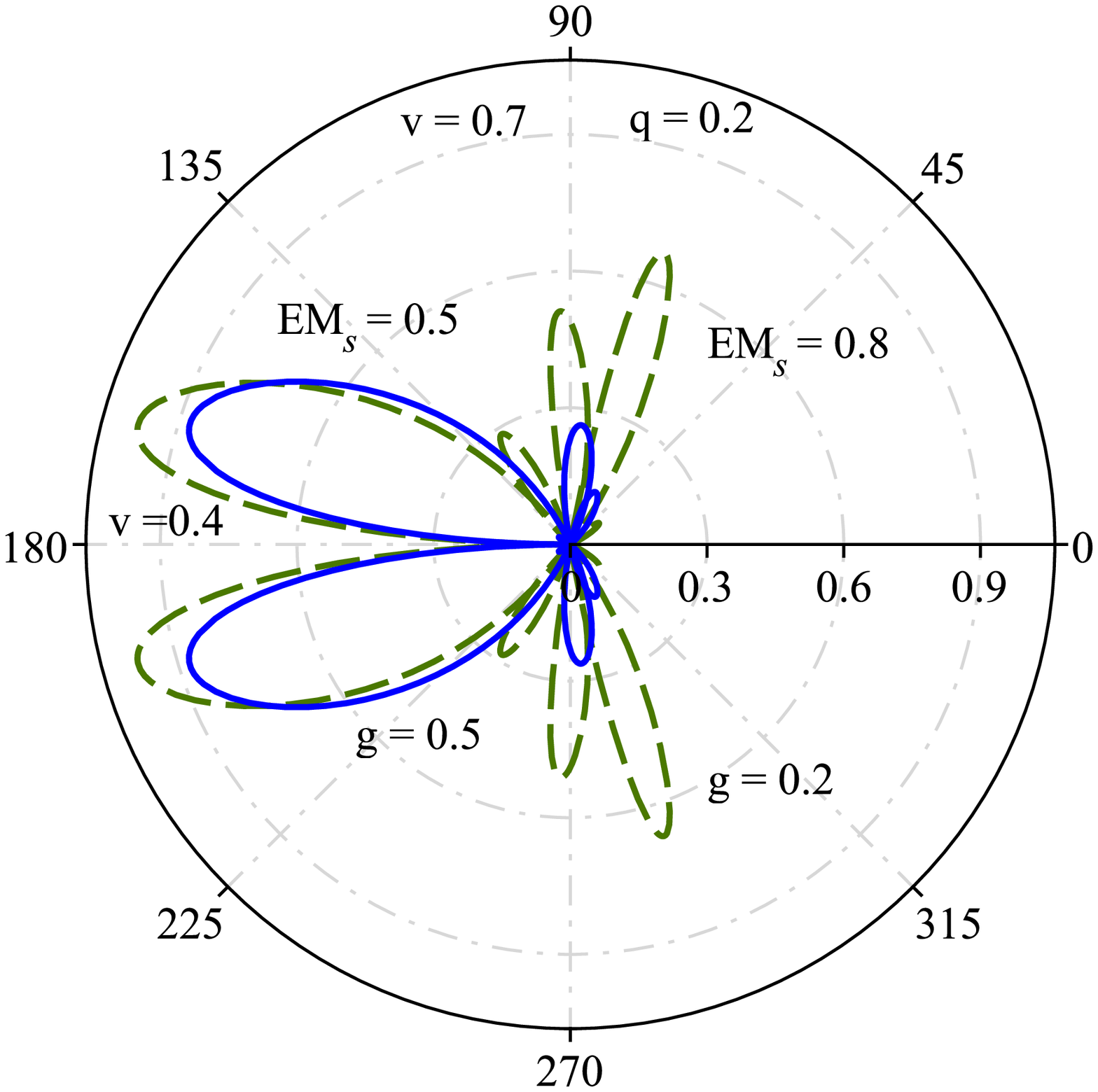}
	\caption{(color online). Polar plots of pure Bardeen (left panel) and Bardeen-class (right panel) polarization $\mathcal{P}(\theta)$ for $0 \leq\theta\leq 2\pi$ showing the alignment of the scattered fermion's spin with a given direction. The Mott polarization (polarization in the direction orthogonal to the scattering plane) can also be observed. }
	\label{fig6}
\end{figure*}

An incident unpolarized beam of massive fermions becomes partially polarized after it gets scattered by the black hole. The induced polarization degree can be computed using the following formula \cite{dolan,rose}
\begin{equation}\label{pwa35}
\mathcal{\vec P}=-i\frac{fg^*-f^*g}{|f|^2 + |g|^2}\vec n
\end{equation}
with $\vec n$ a unit vector in a direction orthogonal to the plane of scattering.

In Fig. \ref{fig5} the dependence of the polarization on the scattering angle is plotted for given values of the parameters $EM,\,v,\,g,\,Q/M$. As can be seen it has a pronounced oscillatory behaviour. From the top panels in Fig.  \ref{fig5} one can observe that if the parameter $EM$ is increasing then the frequency of the oscillations present in the polarization are increasing as well. Now if we  assume that the energy of the fermion is fixed, then this means that the oscillations present in the polarization are more pronounced for black holes with higher masses. Compared with the Schwarzschild polarization (see bottom panels in Fig. \ref{fig5}) the Bardeen and Bardeen-class black hole polarizations have a slightly less oscillatory behaviour. The oscillations present in the polarization can be seen as a consequence of the oscillations present in the glory and spiral scattering.

The alignment of the scattered fermions with the forward on-axis direction can be visualized using polar plots representations of the polarization degree as shown in Fig. \ref{fig6}. One can observe the Mott polarization in the direction orthogonal to the  scattering plane, phenomena also reported before in the literature for Schwarzschild \cite{dolan, sporea1} and Reissner-Nordstr\"om \cite{sporea2} black holes.

\begin{figure*}
	\centering
	\includegraphics[scale=0.44]{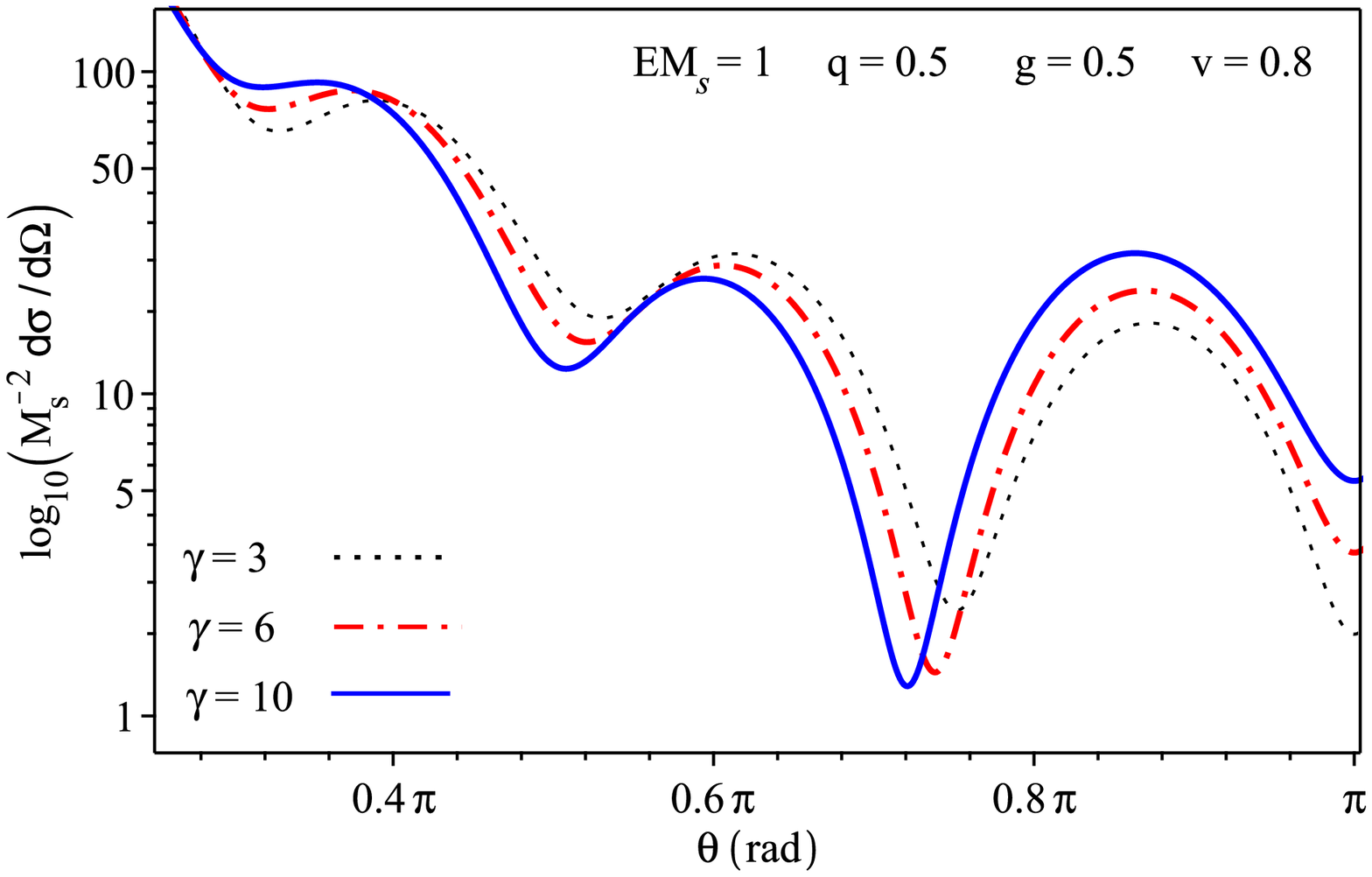}
	\includegraphics[scale=0.44]{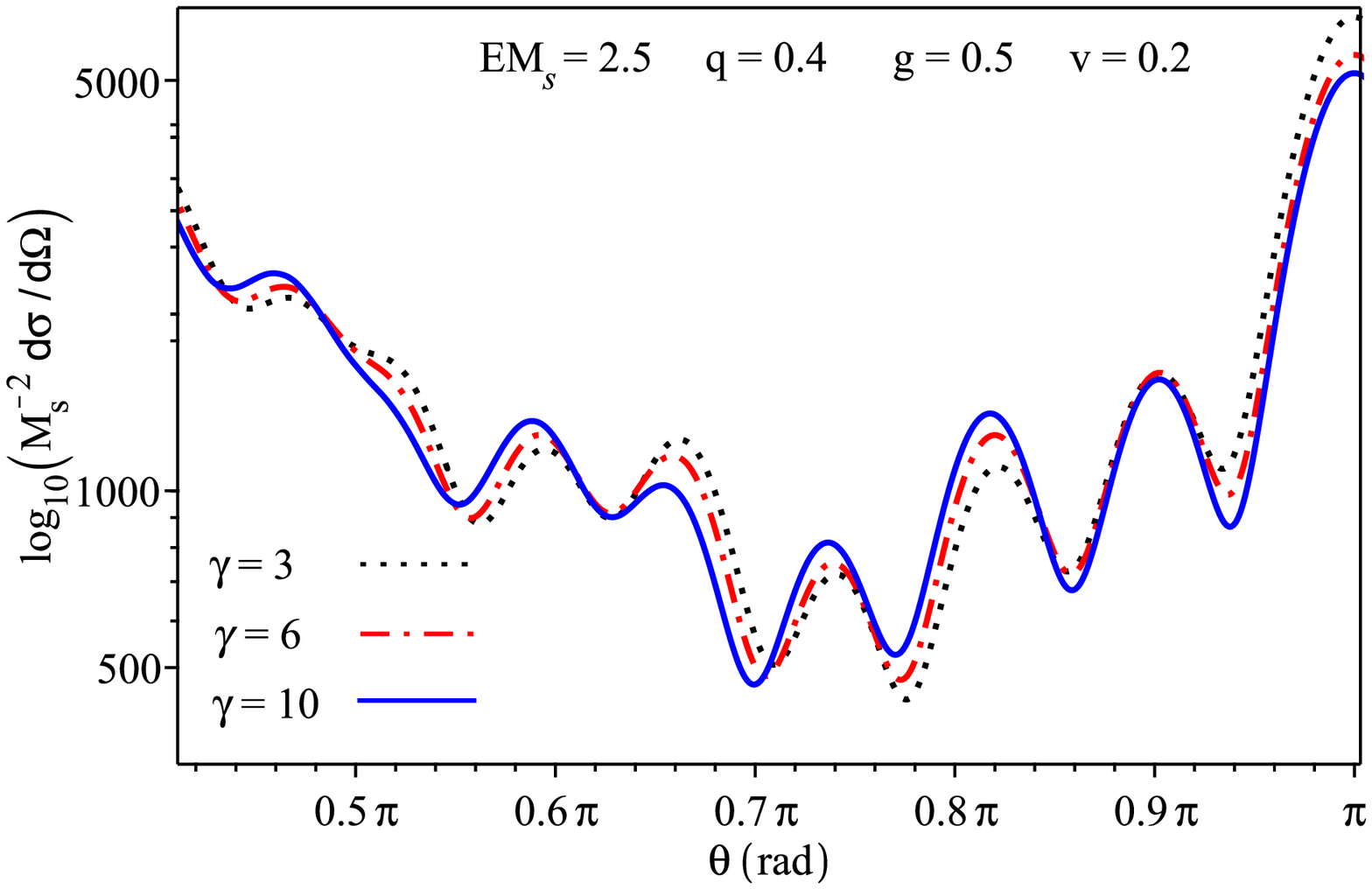}
	\caption{(color online). Comparison of the scattering cross section for a Bardeen-class black hole with different values of the parameter $\gamma$. We notice that the scattering profile keeps the same shape for different values of $\gamma$.   }
	\label{fig4a}
\end{figure*}

In the last Figure, Fig. \ref{fig4a}, the differential scattering cross section is plotted for a Bardeen-class black hole with different values of the parameter $\gamma$. We notice that the scattering pattern keeps the same profile shape and that the difference between the values of $d\sigma/d\Omega$ (computed using different values of $\gamma$) is very small. Moreover, from our analysis results that in the majority of the cases this difference is even smaller than for the examples presented in Fig. \ref{fig4a}. The explanation for this can be traced back to the fact that the difference between the black hole horizon radii $r_+$ obtained for different values of $\gamma$ (while keeping constant all the other parameters) is very small.

\section{Conclusions}\label{sec.final}

In this work we studied the scattering of fermions by a class of Bardeen black holes that include also the original Bardeen regular black hole solution. A partial wave method was used on a set of scattering modes obtained by solving the Dirac equation in the asymptotic region of these black hole geometries. In this way we were able to obtain for the first time analytical phase shifts for fermion scattering by Bardeen regular black holes.  In Section \ref{sec.final} it was shown that the phenomena of glory (scattering in the backward direction) and spiral scattering (oscillations in the scattering intensity) are present. We also saw that an incident unpolarized beam could become partially polarized after the interaction with the black hole.

In Figs. \ref{fig1}-\ref{fig6} besides the parameters $EM$ (that can be associated with a measure of the gravitational coupling) and $v$ (speed of the fermion) we also used the ratios $Q/M$, $q=Q/M_s$ and $g=M/M_s$ to label the figures. The departure of the scattering pattern from the Schwarzschild case becomes significant as $Q/M$ and $g$ increases. For the original Bardeen regular black hole, $Q/M=4/\sqrt{27}$ is the maximum value allowed and it corresponds to the degenerate case, when the two horizons coincide. If the magnetic charge $Q\rightarrow 0$, that also implies $M\rightarrow 0$ and $g\rightarrow 0$, then one recovers the scattering by a Schwarzschild black hole \cite{sporea1}.

As was shown the glory and spiral scattering start to become significant for values of the parameter $EM\sim1$ or bigger (in geometrical units with $G=\hbar=c=1$) due to the fact that for this values the associated wavelength of the incident fermions is of the same order of magnitude as the black hole horizon radius and thus diffraction patterns start to occur. Another feature, that was shown to be present also for fermion scattering by Schwarzschild \cite{dolan,sporea1} and Reissner-Nordstr\"om black holes \cite{sporea2}, is that as the total mass of the black hole is increasing the oscillations present in the scattering intensity become more frequent, meaning that spiral scattering is significantly enhanced with the black hole mass.

\section*{Acknowledgements}

This work was supported by a grant of Ministery of Research and Innovation, CNCS - UEFISCDI, project number PN-III-P1-1.1-PD-2016-0842, within PNCDI III.

\subsection*{Appendix A}\label{appA}

\begin{small}
	
	
The aim of this Appendix is to briefly present how the phase shifts (\ref{pshift}) are obtained.

	\begin{subequations}
		\renewcommand{\theequation}{A\arabic{equation}}
		
		The asymptotic representation of the Whittaker function $M$ for large values of $|z|$ is given by the following formula \cite{NIST}
		
		\begin{equation}\label{A1}
		\begin{split}
		&M_{\kappa,\mu}(z)\sim \frac{\Gamma(1+2\mu)}{\Gamma(\frac{1}{2} 			+\mu-\kappa)}e^{\frac{1}{2}z}z^{-\kappa}\left(1+O(z^{-1})\right) \\
		&+e^{i(\frac{1}{2}+\mu-\kappa)\pi}\frac{\Gamma(1+2\mu)}{\Gamma(\frac{1}{2}+\mu+\kappa)}
		e^{-\frac{1}{2}z}z^{\kappa}\left(1+O(z^{-1})\right)
		\end{split}
		\end{equation}
		valid for $-\frac{1}{2}\pi<{\rm ph \,}z<\frac{3}{2}\pi$.
	
		For obtaining the phase shifts (\ref{pshift}) the condition $C_2=0$ must be imposed in eq. (\ref{sol10}) in order to obtain the correct Newtonian phase shifts for large values of angular momenta. As an observation this condition also selects the spinors that are regular in $x=0$ due to the regularity of the function $M_{\rho_{\pm},s}(2i\nu x^2)=(2i\nu x^2)^{s+\frac{1}{2}}[1+O(x^2)]$. More arguments for why one must take $C_2=0$ can be found in Appendix C of our previous paper \cite{sporea1}.
		
		Making use of Eq. (\ref{A1}) on the $\hat f^\pm(x)$ functions and by observing that for $\hat f^+$ the dominant term is the first one in (\ref{A1}), while for $\hat f^-$ the second term is dominant, then one obtains the following asymptotic expressions
		\begin{equation}\label{A2}
		\openup 2\jot
		\begin{split}
		&\hat f^+(x)\to C_1 e^{-\frac{1}{2}\pi \alpha}\frac{\Gamma(2s+1)}{\Gamma(1+s+i\alpha)}\,e^{i[\nu x^2+\alpha\ln(2\nu x^2)]}\\
		&\hat f^-(x)\to \frac{(s-i\alpha)(\kappa+i\lambda)}{\kappa^2+\lambda^2}C_1e^{-\frac{1}{2}\pi \alpha}\frac{\Gamma(2s+1)}{\Gamma(1+s-i\alpha)}\times \\
		&\qquad\qquad\qquad \times e^{-i[\nu x^2+\pi s+\alpha\ln(2\nu x^2)]}
		\end{split}
		\end{equation}
		
		The last step is to compute the argument of the trigonometric functions (\ref{arg}) as $\frac{1}{2}arg\left(\frac{\hat f^+}{\hat f^-}\right)$ from which the expression for $e^{2i\delta_{\kappa}}$ will result.

	\end{subequations}
	
\end{small}

\end{multicols}

\vspace{-1mm}
\centerline{\rule{80mm}{0.1pt}}
\vspace{2mm}

\begin{multicols}{2}

\end{multicols}

\clearpage
\end{CJK*}
\end{document}